\documentclass[iop]{emulateapj}

\usepackage{times,color,natbib,url,amsmath}
\usepackage{graphicx,float,psfrag}
\usepackage{tabularx}
\usepackage{latexsym,amsmath,amssymb}
\usepackage{natbib}
\usepackage{verbatim}
\usepackage{multirow}

\bibpunct{(}{)}{;}{a}{}{,}

\def \george    {G. Younes}

\def \chryssa   {C. Kouveliotou}
\def \pavlov {G.~G. Pavlov}
\def \oleg {O. Kargaltsev}
\def \stefanie {S. Wachter}
\def \ersin {E. G\"o\u{g}\"u\c{s}}

\newcommand {\xmm} {\textsl{XMM-Newton}}
\newcommand {\chandra} {\textsl{Chandra}}
\newcommand {\swift} {\textsl{Swift}}
\newcommand {\rxte} {\textsl{RXTE}}
\newcommand {\fermi} {\textsl{Fermi}}

\def \rsun {\ifmmode$R$_{\odot}\else R$_{\odot}$}

\def \hcm {\hbox {\ifmmode $ atoms cm$^{-2}\else atoms cm$^{-2}$\fi}}

\def\approxgt{\mathrel{\hbox{\rlap{\lower.55ex \hbox {$\sim$}}
        \kern-.3em \raise.4ex \hbox{$>$}}}}
\def\approxlt{\mathrel{\hbox{\rlap{\lower.55ex \hbox {$\sim$}}
        \kern-.3em \raise.4ex \hbox{$<$}}}}

\def \arcmin {\hbox{$^\prime$}}
\def \arcsec {\hbox{$^{\prime\prime}$}}

\def \src {Swift~J1834.9--0846}
\def \hessj{HESS~J1834--087}

\begin{document}

\title{\xmm\ view of \src\ and its Magnetar Wind Nebula}

%
%
\author{\george$^{1,2}$, \chryssa$^{3,2}$, \oleg$^4$, \pavlov$^{5,6}$, \ersin$^7$, and
  \stefanie$^8$}
\affil{
 $^1$ Universities Space Research Association, 6767 Old Madison Pike NW, Suite 450, Huntsville, AL 35806, USA \\
 $^2$ NSSTC, 320 Sparkman Drive, Huntsville, AL 35805, USA \\
 $^3$ Astrophysics Office, ZP12, NASA/Marshall Space Flight Center, Huntsville, AL 35812, USA \\
 $^4$ Dept.\ of Astronomy, University of Florida, Bryant Space Science Center, Gainesville, FL 32611, USA \\
 $^5$ Pennsylvania State University, 525 Davey Lab, University Park, PA 16802, USA \\
 $^6$ St.-Petersburg State Polytechnical University, Polytekhnicheskaya ul.\ 29, 195251, St.-Petersburg, Russia \\
 $^7$ Sabanc\i University, Faculty of Engineering and Natural Sciences, Orhanl\i$-$Tuzla, \.{I}stanbul 34956, Turkey \\
 $^8$ Infrared Processing and Analysis Center, California Institute of Technology, Pasadena, CA 91125, USA 
}

\date{}
%
%

\begin{abstract}

We report on the analysis of two \xmm\ observations of the recently
discovered soft gamma repeater \src, taken in September 2005 and one
month after the source went into outburst on 2011 August 7. We
performed timing and spectral analyses on the point source as well as
on the extended emission. We find that the source period is consistent
with an extrapolation of the \chandra\ ephemeris reported earlier and
the spectral properties remained constant. The source luminosity
decreased to a level of $1.6\times10^{34}$ erg~s$^{-1}$ following a
decay trend of $ \propto t^{-0.5}$. Our spatial analysis of the source
environment revealed the presence of two extended emission regions
around the source. The first (Region A) is a symmetric ring around the
point source, starting at 25\arcsec\ and extending to
$\sim$50\arcsec. We  argue that Region A is a dust scattering
halo. The second (Region B) has an asymmetrical shape extending
between 50\arcsec\ and  150\arcsec, and is detected both in the pre-
and post-outburst data. We argue that this region is a possible
magnetar wind nebula (MWN). The X-ray efficiency of the MWN with
respect to the rotation energy loss is substantially higher than those
of rotation powered pulsars: $\eta_{\rm X}\equiv L_{\rm
  MWN,0.5-8~keV}/\dot{E}_{\rm rot}\approx0.7$. The higher efficiency
points to a different energy source for the MWN of \src, most likely
bursting activity of the magnetar, powered by its high magnetic field,
$B=1.4\times10^{14}$ G.

\end{abstract}

\section{Introduction}
\label{Sec:Intro}

Soft Gamma Repeaters (SGRs) and Anomalous X-ray Pulsars (AXPs) are two
empirical classes of objects widely accepted to comprise the magnetar
population, i.e., isolated neutron stars with ultra-strong magnetic
fields ($B\gtrsim10^{14}-10^{15}$~G). Their existence was predicted
theoretically in 1992 \citep{duncan92ApJmagnetars,
  paczynski92AcA:magnetars}, but was only confirmed in 1998 with {\it
  RXTE} observations (\citealt{kouv98Natur:1806,
  kouveliotou99ApJ:1900}; for detailed magnetar reviews please refer 
to \citealt{woods06csxs:magnetars}, \citealt{
  mereghetti08AARv:magentars}). SGRs and AXPs share many
characteristics such as long spin periods (2-12 s) and large spin-down
rates that imply very high surface dipole magnetic fields of
$10^{14}-10^{15}$~G. They are all persistent X-ray emitters with
luminosities significantly larger than those expected from rotational
energy losses; instead the magnetar X-ray emission is attributed to
the  decay of their powerful magnetic fields and sub-surface heating
\citep{thompson96ApJ:magnetar}. Magnetars enter active episodes during
which they emit short (0.1\,s) bursts of hard X-/soft $\gamma-$rays
with luminosities ranging from $10^{37}$ to 10$^{41}$~erg~s$^{-1}$;
very rarely, they emit Giant Flares (GFs) that 
last several minutes with luminosities
$\gtrsim10^{46}$~erg~s$^{-1}$. The typical magnetar bursts are
attributed to neutron star crust quakes caused by the evolving
magnetic field under its surface \citep{thompson95MNRAS:GF}. 

\begin{figure}[th]
\begin{center}
\includegraphics[angle=0,width=0.48\textwidth]{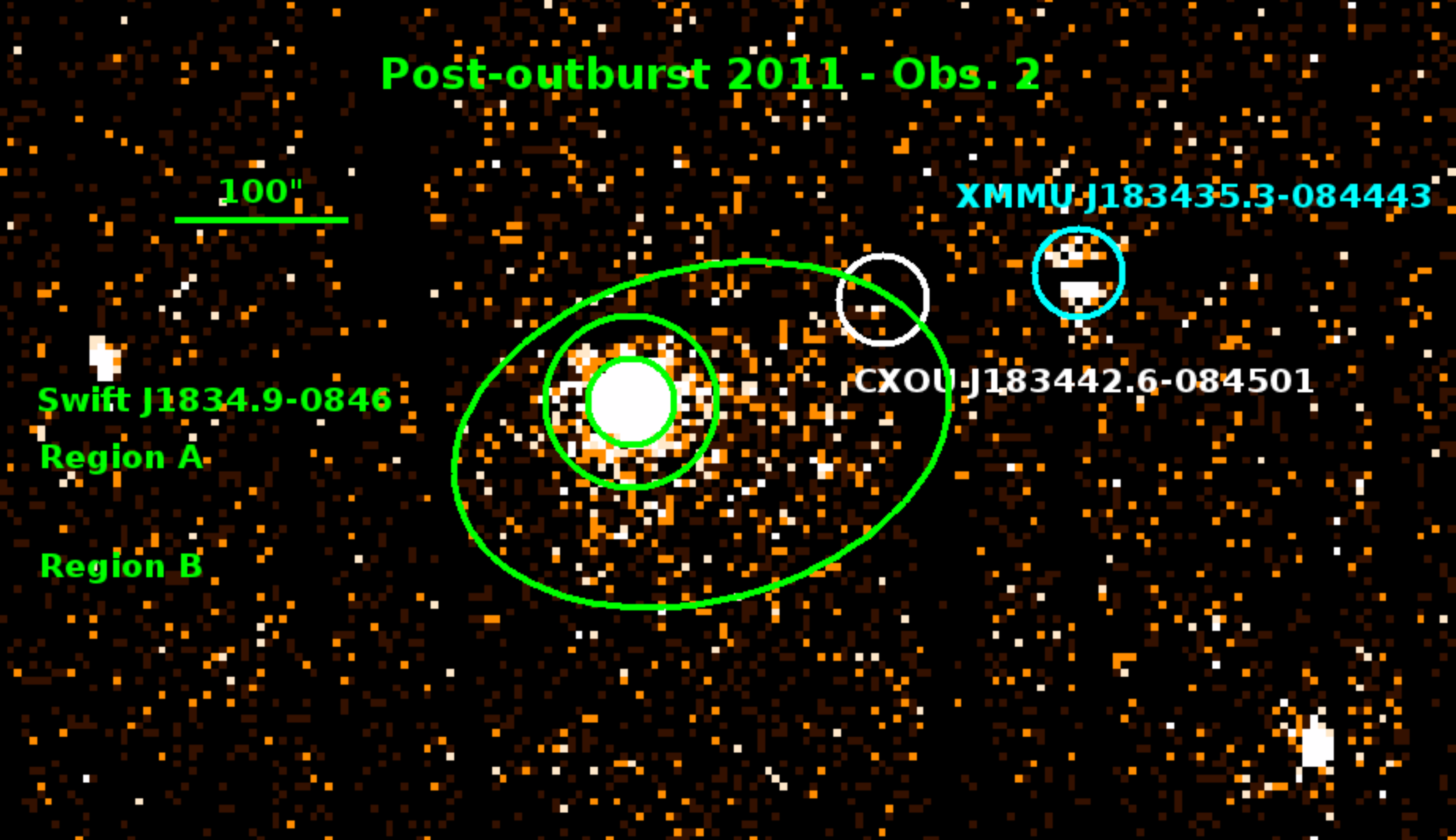}\\
\includegraphics[angle=0,width=0.48\textwidth]{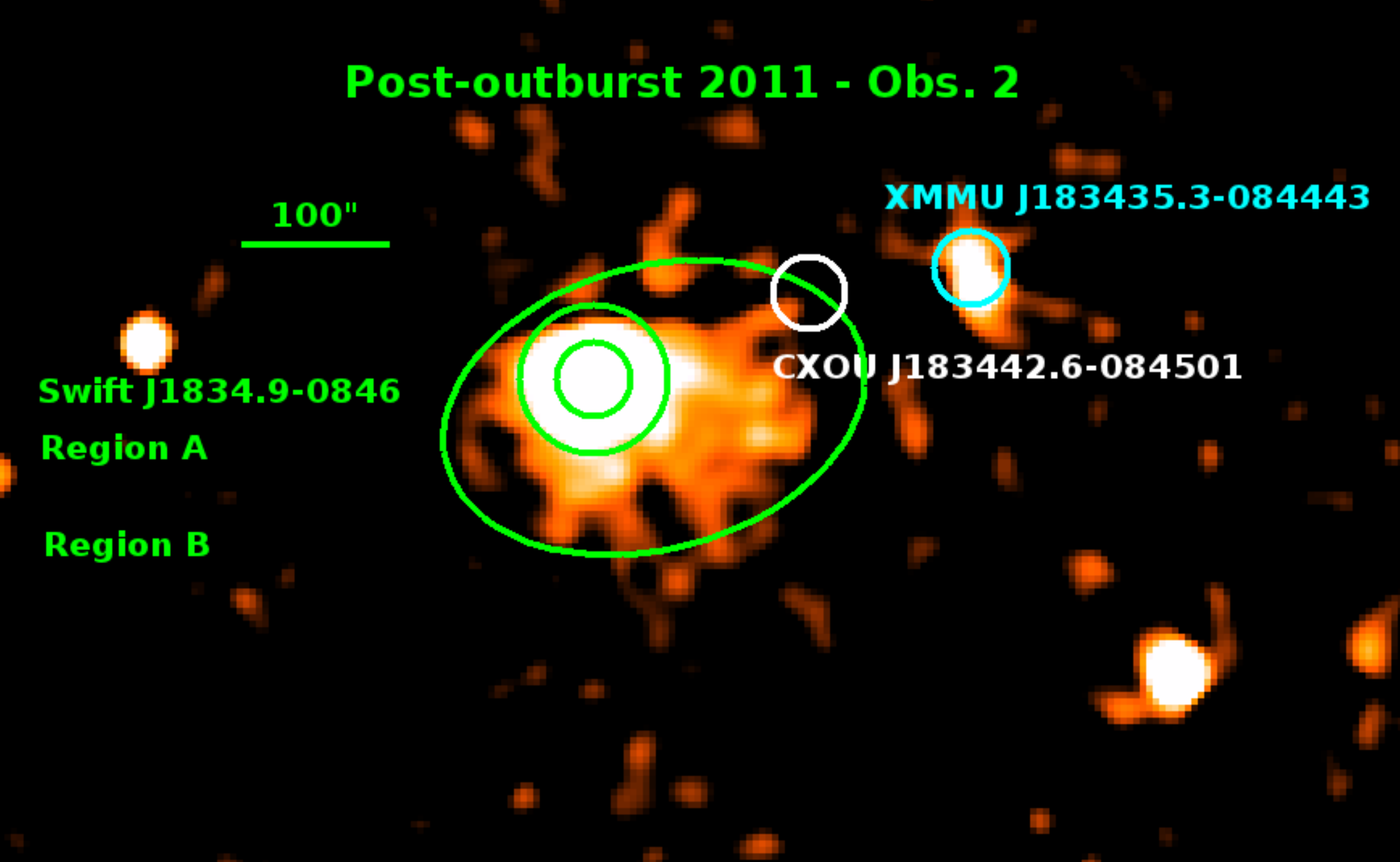}\\
\includegraphics[angle=0,width=0.48\textwidth]{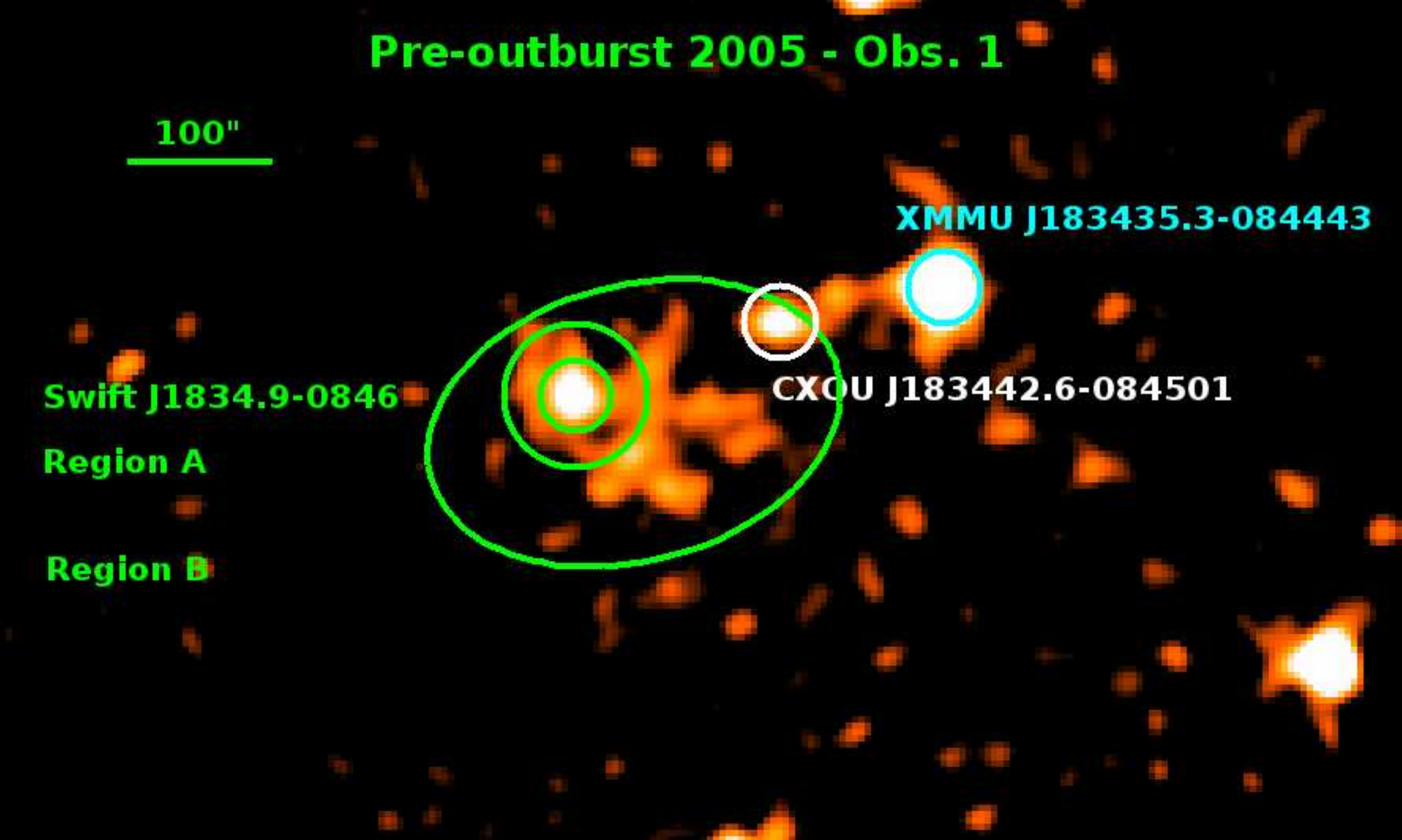}
\caption{Post-outburst {\it XMM-Newton} EPIC-PN observation of \src\ in 2011
  (obs.~2, upper and middle panels) and pre-outburst 2005 EPIC MOS1+MOS2 observation
  (obs.~1, bottom panel). The  middle and bottom images are Gaussian
  smoothed with a FWHM of 5.0 pixels (20\arcsec). The smallest green
  circle with a 25\arcsec-radius represents the \src\ point-source
  emission. The annulus with 25\arcsec$\le r\le$50\arcsec\ represents the symmetrical extended emission around the
  point source (region A). The ellipse of major (minor) axis of 145\arcsec\ (95\arcsec) encloses the asymmetrical extended emission
  around \src\ (region B). Other sources in the field are labeled. North is up and west is right.}
\label{imagxray}
\end{center}
\end{figure}

An interesting question in the magnetar field is their evolutionary link, if any, to their less magnetically-powerful counterparts, rotation powered pulsars (RPPs). The latter sources are known to produce particle outflows, often resulting in spectacular Pulsar Wind Nebulae (PWNe, see \citealt{kargaltsev08PWN} for a review) of which the Crab is the most famous example \citep{weisskopf00ApJ}. The PWN X-ray emission is due to synchrotron radiation from the shocked relativistic outflow of electrons and positrons produced by the pulsar.  Magnetars are also expected to produce particle outflows, either steady or released during outbursts accompanying bright bursts or GFs \citep{thompson98PhRvD:mag, harding99ApJ:mag,tong12arXiv1205:mwn}. The GF of 2004 December 27 from SGR\,J$1806-20$ released at least $4\times10^{43}$ ergs of energy in the form of magnetic fields and relativistic particles \citep{gaensler05Natur:1806}. Given the strong magnetic fields associated with this class of neutron stars, the idea, therefore, of a Magnetar Wind Nebula (MWN) seems very plausible.

Only a few claims have been made so far for the detection of a nebula
around a magnetar. The first one was the radio nebula around
SGR~J$1806-20$ \citep{kulkarni94Natur:1806}, which was shown later to
be enshrouding a Luminous Blue Variable star, unrelated to the SGR
\citep{hurley99ApJsgr1806}. Elongated and expanding radio emission was
unambiguously identified following the GF of SGR~J$1806-20$
\citep{gaensler05Natur:1806,gelfand05ApJ:sgr1806}, most likely
associated with jets produced by the flare. A variable radio
  source indicating particle outflow was also seen after the giant
  flare of SGR~1900+14 \citep{frail99Natur:sgr1900}. Recently,
\citet{rea09ApJpwn}, \citet[][see also \citealt{
  gonzalez03ApJ:PSR1119}]{safiharb08ApJ:PSR1119} and
\citet{vink09ApJ:PWN} reported the discovery of unusual extended
emission around three high $B-$field sources, a Rotating Radio
Transient, RRAT~J1819$-$1458, a high-B pulsar PSR J1119-6127, and a
magnetar 1E~$1547.0-5408$ (SGR~J$1550-5418$) respectively. The latter case
was shown to be a halo on the basis of correlated flux variations in
the extended emission and the magnetar
\citep{olausen11ApJ:sgr1550}. In summary, to date there is no 
unambiguous evidence for the existence  of a PWN/MWN around a
magnetar. Confirmed detections of MWNe would reconcile observations
with theoretical predictions of their existence and would shed light
on the nature of magnetar outflows and the environmental properties of
magnetars. 

\src\ is the last in a long line of magnetar discoveries during the
last three years, owing to the synergy between NASA's three
observatories, \swift, \rxte, and \fermi. It was discovered on 2011
August 7, when it triggered the \swift/Burst Alert Telescope (BAT) and
the Fermi/Gamma-ray Burst Monitor (GBM) with a soft, short burst
\citep{delia11GCN1834,guiriec11GCN:1834}. The magnetar nature of \src\
was established with {\it RXTE}/PCA and \chandra\ Target of
Opportunity observations, which revealed a coherent X-ray pulsation at
a spin period $P=2.482295$~s
\citep{gogus11ATel:1834,gogus11ATel:1834B}, and a spin-down rate
$\dot{\nu}=-1.3(2)\times10^{-12}$~Hz~s$^{-1}$ \citep{kuiper11:1834},
implying a dipole surface magnetic field $B=1.4\times10^{14}$~G, and a
spin-down age and energy loss rate $\tau=4.9$~kyr and $\dot{E}_{\rm
  rot}=2.1\times10^{34}$~erg~s$^{-1}$, respectively.

\citet[][K+12 hereinafter]{kargaltsev12apj:1834} studied the
spatial, temporal, and spectral properties of \src\ using the
available \swift, \rxte, and \chandra\ post-outburst observations, and
one \chandra\ pre-outburst observation taken in June 2009. The
persistent X-ray light curve of the source, spanning 48 days after the
first burst, showed that the $2-10$~keV flux decayed steadily as a
power-law with index $\alpha=0.53\pm0.07$ ($F\propto
t^{-\alpha}$). The source spectrum ($2-10$\,keV) was well fit with
either an absorbed power-law with a photon index $\Gamma\approx3.5\pm0.5$
or an absorbed blackbody with a temperature
$kT=1.1\pm0.1$~keV, and an emitting area radius of 0.26 km (assuming a
source distance of 4 kpc, see below). The hydrogen column density was
of the order of $10^{23}$~cm$^{-2}$, depending on the model
spectrum. Finally, K+12 reported the presence of an extended emission
up to a radius of 10\arcsec\ from the center of the source, most
likely a dust scattering halo, considering the large absorption toward
the source position. However, an even more extended emission, with
radius $>$30\arcsec, was detected in the 2009 pre-outburst \chandra\
observation. The asymmetrical shape of this emission,
northeast-southwest of the source, poses a challenge to the dust
scattering halo interpretation, especially since this extended
component was detected while the point source was not seen down to a
limit of $10^{-15}$~erg~cm$^{-2}$~s$^{-1}$.

Here we report the analysis of two \xmm\ observations of \src, taken
in September 2005 and September 2011 (one month after the source
outburst), with emphasis on the analysis of the environment around the
source. Section 2 describes the observations and data reduction
techniques. We present our results of the spatial, timing and spectral
analysis in Section~3. We discuss the spectral and temporal results of
\src\ and the implication of our extended emission analysis in the
context of Magnetar Wind Nebula (MWN) in Section~4. Given a
  plausible association between Swift J1834.9–0846 and the SNR W41, we
  will assume that both are at the same distance ($\sim4$ kpc,
  \citealt{tian07ApJ:1834,leahy08AJ:w41}; K+12) throughout the paper.

\section{\xmm\ observations and data reduction}
\label{sec:obs}

\begin{figure*}[th]
\begin{center}
\includegraphics[angle=0,width=0.48\textwidth]{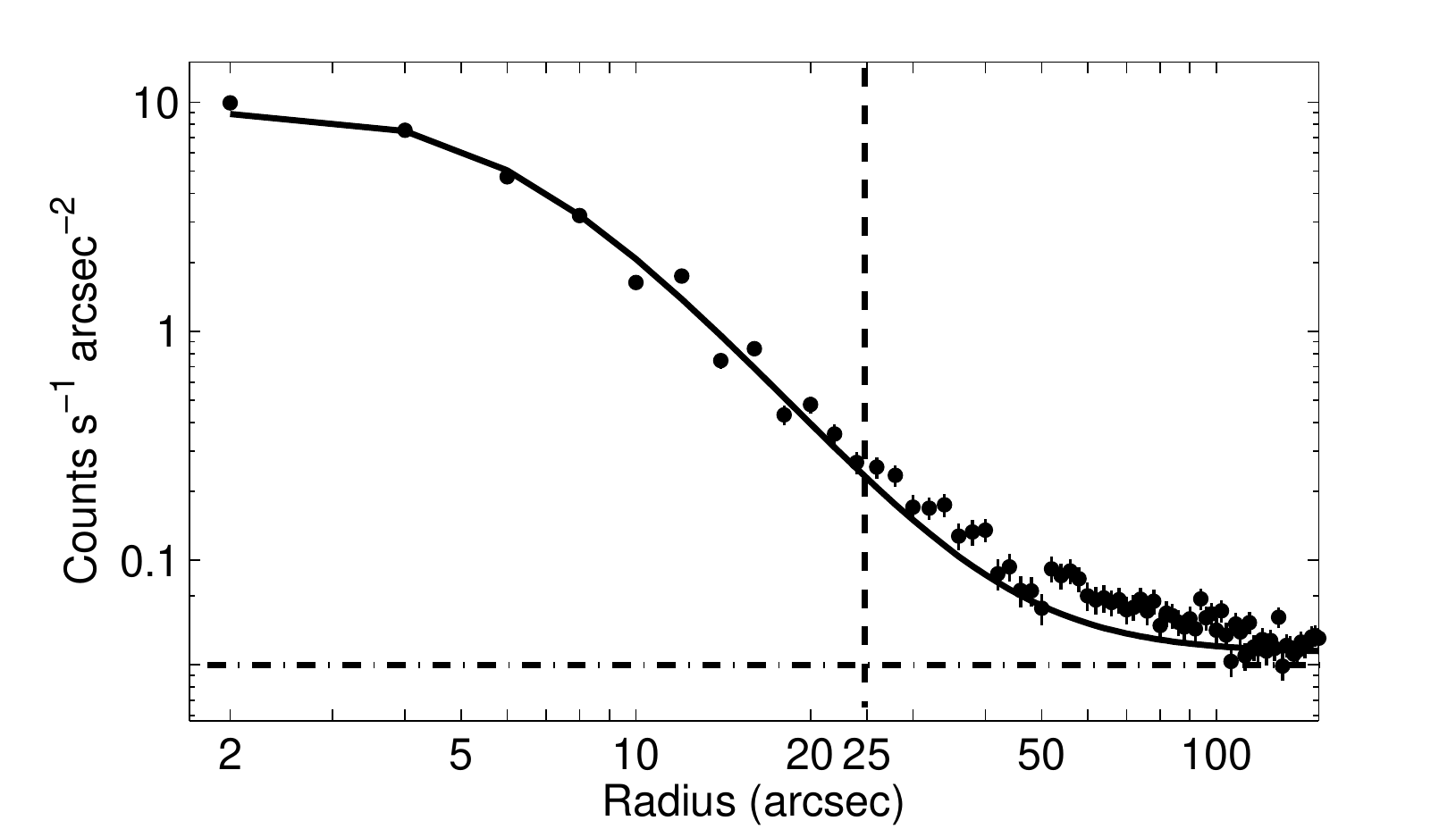}
\includegraphics[angle=0,width=0.48\textwidth]{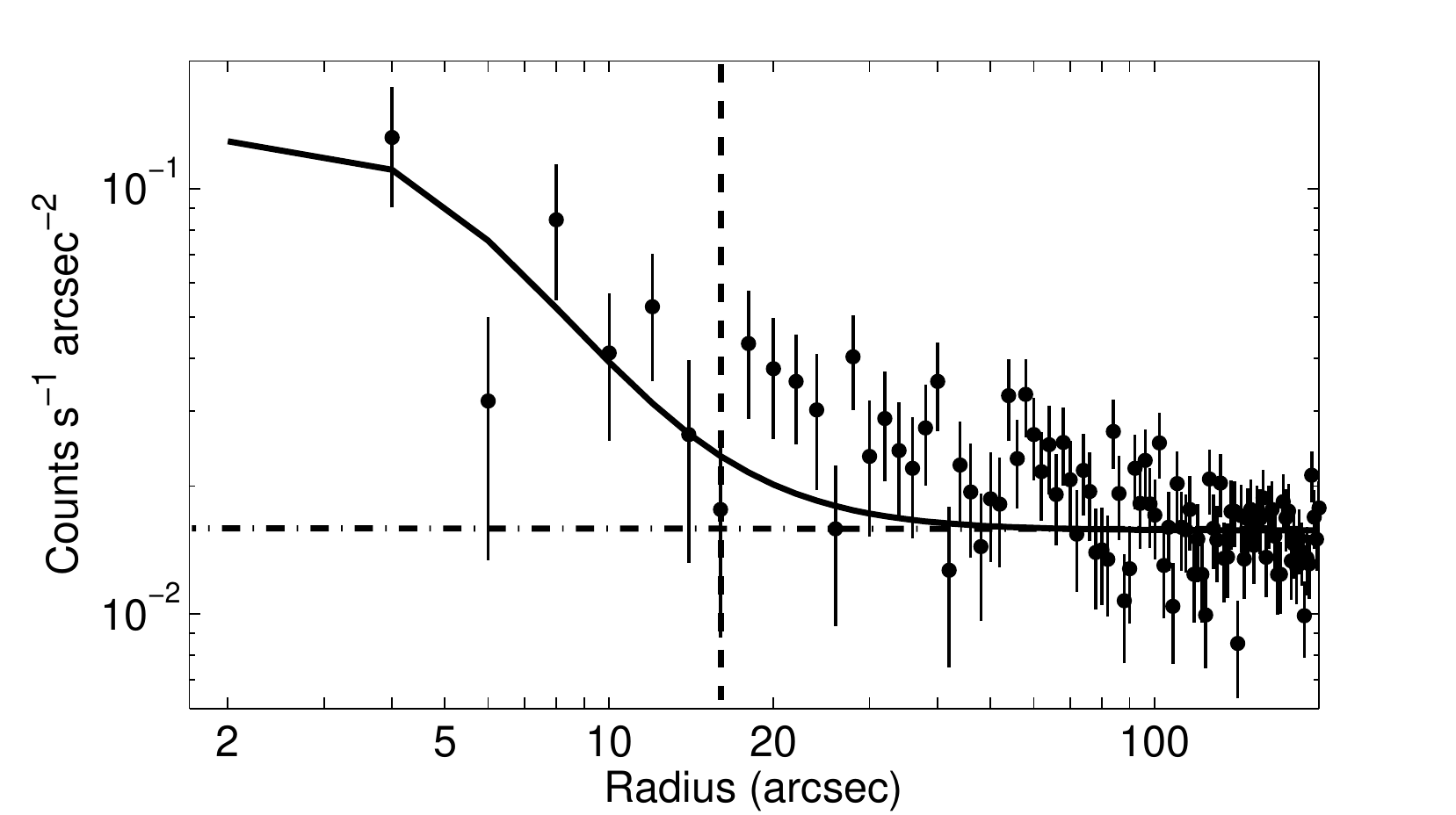}
\caption{Radial profile of the X-ray emission ($1.5-8$~keV) of \src\
  using the \xmm\ data from the post-outburst Obs.~2 (PN, {\sl left
    panel}) and the pre-outburst Obs.~1 (MOS 1+2, {\sl right
    panel}). The black  solid line represents the best-fit PSF for
  each camera. Extended emission is clearly seen beyond
  $\sim$20\arcsec\ and $\sim$15\arcsec\ in obs.~2 and obs.~1,
  respectively.}
\label{radprof-fig}
\end{center}
\end{figure*}

\begin{figure*}[!th]
\begin{center}
\includegraphics[angle=0,width=0.45\textwidth]{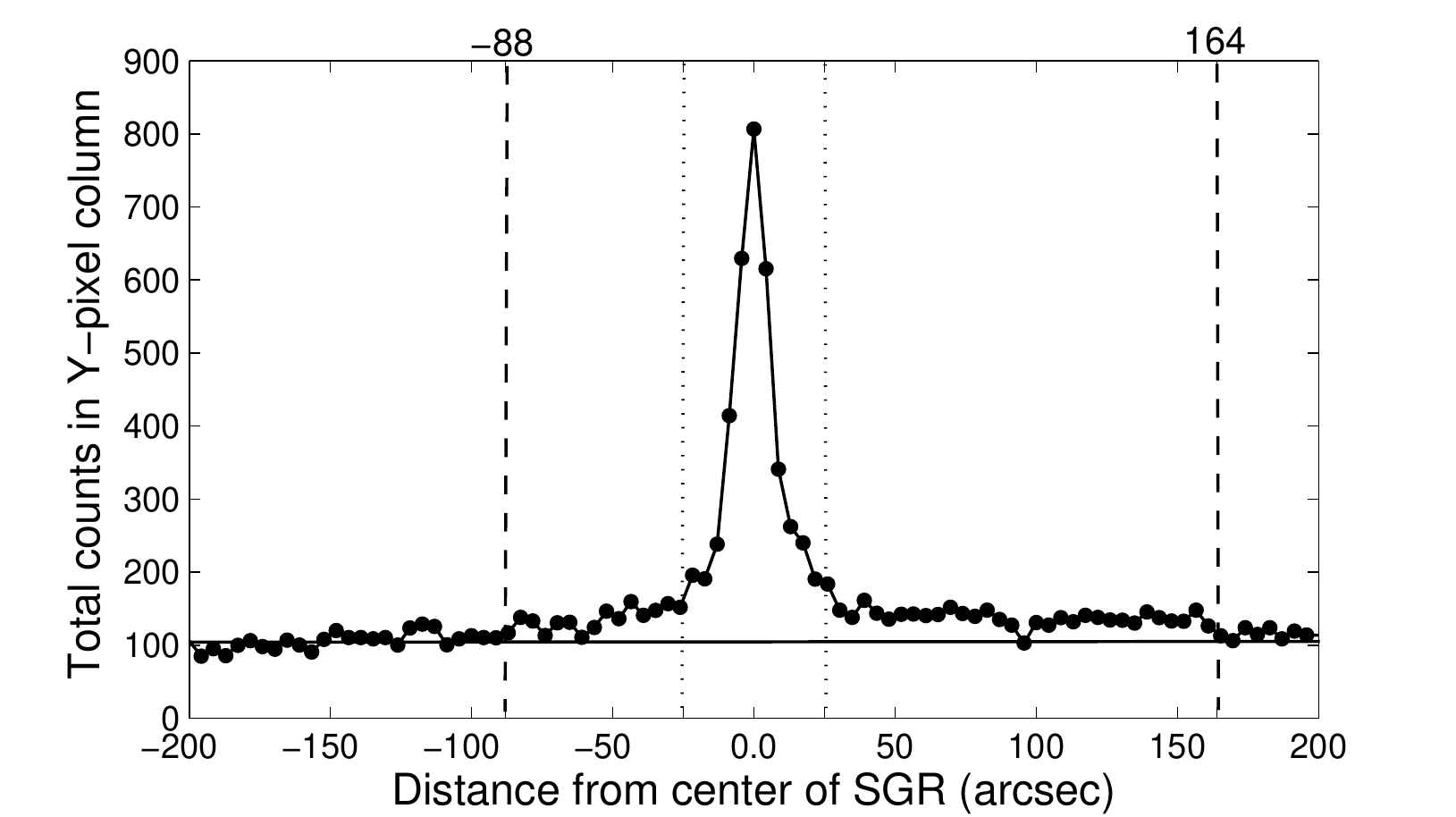}
\includegraphics[angle=0,width=0.45\textwidth]{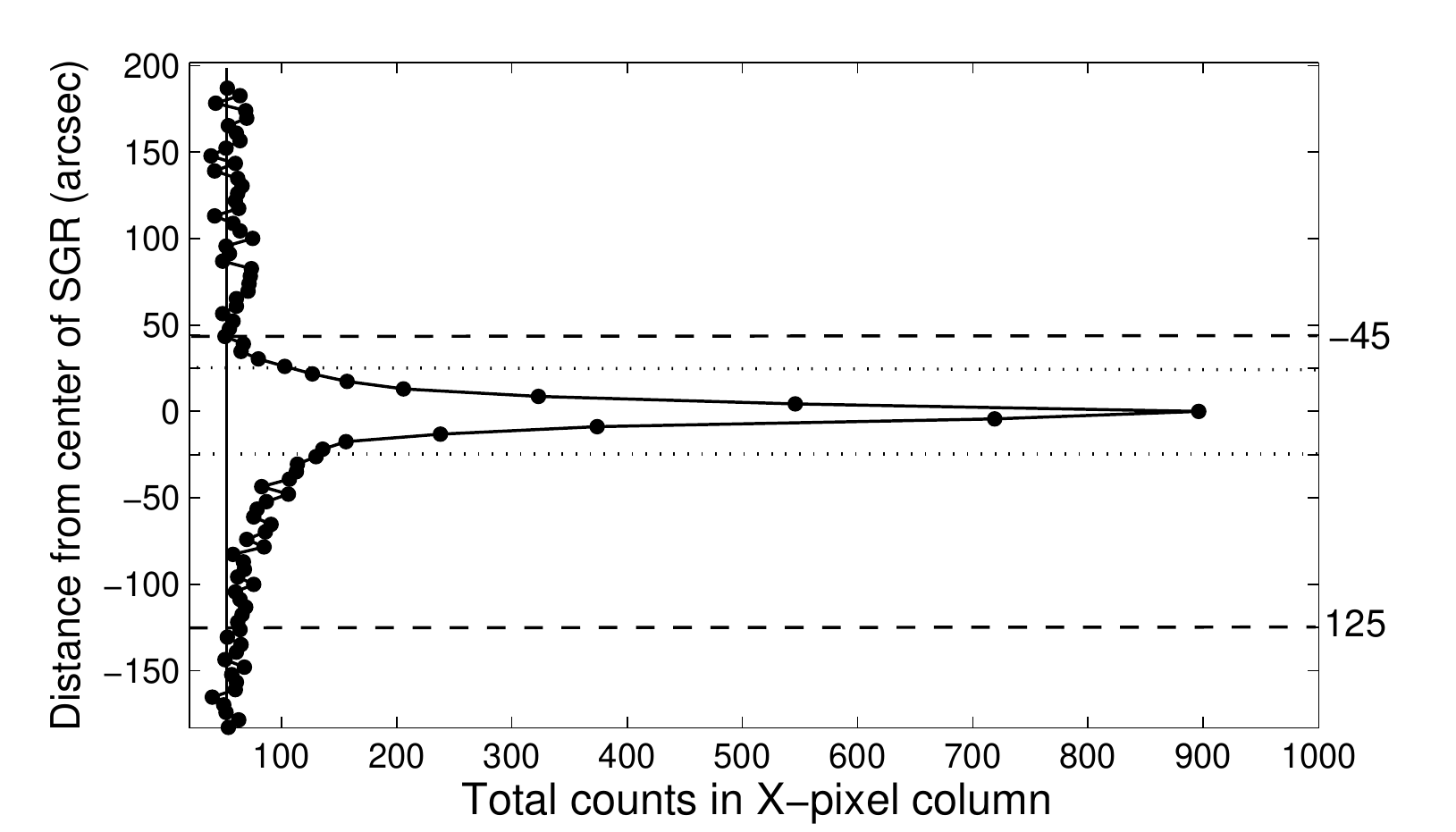}
\caption{{\sl Left column.} Projected total counts of a pixel column in the
  Y-direction (north-south, left panel) and X-direction (east-west,
  right panel) in a rectangular region around
  \src. The dotted lines delimit a 25\arcsec\ circular region
  around the SGR. The dashed lines represent the extent of the asymmetrical extended
  emission.}
\label{extent}
\end{center}
\end{figure*}

The field of  the newly discovered magnetar, \src, was observed twice
with \xmm. The  first  observation (obs.ID 0302560301, obs.~1
hereafter, PI Gerd Puehlhofer), taken in 2005 September  for an
exposure time  of about 20~ks,  was intended to image the \hessj\
field  in which \src\ lies. During this observation, \src\ was
  $\sim2$\arcmin\ off-axis from the nominal on-axis position, which is
  small enough not to cause substantial vignetting. The EPIC-PN  and
MOS detectors were operating in Prime Full Frame mode using the medium
filter. Data from all three EPIC instruments were analyzed in the past
(EPIC-PN, \citealt{tian07ApJ:1834}; EPIC-MOS, \citealt{
  mukherjee09ApJ:1834}).  We re-analyzed this observation  to look for
an extended emission at the position of \src.

The second  \xmm\ observation (obs.ID 0679380201, obs.\,2 hereafter)
was a TOO (PI Norbert Schartel) taken on 2011 September 17 for an
exposure of about 24~ks, with \src\ being at the aimpoint  of the
three  EPIC detectors.  The EPIC-PN detector was  operating  in Prime
Full Frame mode using the medium filter. The EPIC-MOS detectors, on
the other hand, were operating in Small Window mode.

The two observations were reduced and analyzed in a homogeneous manner
using the Science Analysis System (SAS) version 11.0.0 and FTOOLS
version  6.11.1. Data  were selected using event patterns  0--4 and
0--12 for  PN and MOS, respectively, during  only good X-ray events
(``FLAG==0''). We excluded intervals of enhanced particle background
during Obs.~1, resulting in an effective exposure time of $\sim14$~ks
in the MOS cameras. Response matrices were generated using the
  task {\sl rmfgen}. These responses were spatially averaged using a
  PSF model for point-like sources and a flat uniform flux
  distribution for extended sources.

Background events for point-like sources were extracted from a
source-free region with the same size as the source on the same
CCD. We followed the same procedure for the background extraction of
extended sources since they only cover a small region in the sky with
a size of 2-3\arcmin\ (see Section~\ref{spa-ana}).

For point-like sources the background spectrum was directly
subtracted from the source spectrum. Such a method corrects for both
the instrumental and the cosmic X-ray background
simultaneously. Since our extended sources are not very large (see
Section ~\ref{spa-ana}) one can expect that same method would work
reasonably well for their spectra. However, to ensure that the
background contribution is accurately accounted for, we also tried a
more rigorous background-estimate procedure, where we first modeled
the background spectrum and then included the background
contribution as an additional model component while fitting the
source spectrum.

We used the Extended Science Analysis Software (ESAS)
  package\footnote{http://xmm.esac.esa.int/sas/current/doc/esas/index.html} 
for the purpose of background modeling. First, the instrumental
background is extracted from the CCDs where our extended emission
lies, using the filter-wheel closed data, i.e., derived from
observations where the filter wheel is in the closed position. We
correct both the background and the source spectra for the
instrumental background. Then, we fit the resulting background spectrum with a
combination of two thermal components and an absorbed power-law. We
froze the temperature of one of the thermal components to 0.1 keV
assuming emission from the local hot bubble. The temperature of the
second thermal model, which represents the emission from the
interstellar/intergalactic medium, was left free to vary \citep{
  snowden04ApJ:xmmbackgr, snowden08AA:xmmbackgr}. The 
absorption in the power-law was frozen to the Galactic value towards
\src, $N_{\rm H}=1.63\times10^{22}$~cm$^{-2}$, and the photon index of
the power-law was frozen to 1.5 assuming unresolved AGN contribution
(e.g., distant quasars and/or nearby low luminosity AGN,
\citealt{porquet04aa:pgquasar, sazonov08AA:xrb,
  younes11AA:liner1sXray}). We also added a Gaussian emission line
with a centroid energy of 1.5~keV to model the instrumental EPIC-PN
Al~K$\alpha$ line. The model fit to the background spectrum was good,
with $\chi_{\nu}^2=1.3$ for 42 d.o.f. The temperature of the thermal
component is $kT\approx1.0$~keV, a reasonable value for the
intergalactic medium X-ray emission. Finally, we fit the extended
emission spectra with an absorbed power-law, including the background
best-fit model.

The best fit parameters to our extended sources spectra using the
  two background-estimation methods, i.e., directly or through
  modeling, were in very good agreement within the error bars at the 1
  sigma level (Table~1). Hence, in the following the background for
  extended sources was estimated directly, as usually done for
  point-like sources.

\section{Results}

\subsection{Spatial analysis}
\label{spa-ana}

The X-ray images ($1.5-8$ keV) of \src\  are shown in
Figure~\ref{imagxray} for obs.~1 (MOS 1+2 cameras, lower panel) and
obs.~2 (PN camera, upper and middle panels)\footnote{During obs.~1
  \src\ lies on a CCD gap in the PN camera and these data were not
  used; obs.~2 used MOS cameras in Small Window mode.}. The middle and
lower panels are smoothed with a Gaussian of FWHM$\simeq20$\arcsec\ to
accentuate the extended emission.

We extracted the radial profile from a set of circular annuli
centered at the position of \src\ using the MOS 1+2 and PN cameras for
obs.~1 and obs.~2, respectively (Figure \ref{radprof-fig}). These
  radial profiles were then fit by re-normalizing a \xmm\ PSF template
  (to have similar number of counts at the core) and adding a constant
  background (dot-dashed line). This
  PSF template, given as an XMM-Newton calibration file
  (XRT3\_XPSF\_0013.CCF), is the best fit King function 
  \citep{king66AJ:radprof} to the radial profile of many bright point
  sources observed with the EPIC  cameras. The rms values  of the PSF
  fit to our radial profiles are 0.10 and 0.35 for  obs.~1 and obs.~2,
  respectively, indicating that a PSF  alone is not
  sufficient to explain the observed source radial profiles, and  that an
  excess emission is present. Indeed, extended emission is clearly
visible in  both observations, starting at around 15\arcsec\ and
25\arcsec\ for obs.~1 and obs.~2, respectively. The extent of this
emission is larger and more obvious in  obs.~2, stretching out to 
$r\gtrsim$150\arcsec. The emission in obs.~1 is detected up to
$r\approx$70\arcsec.

It is clear from Figure~\ref{imagxray} (middle panel) that the
extended emission around \src\ becomes asymmetrical in shape at
$r\approx$50\arcsec. We quantified the asymmetrical shape of this
extended emission in obs.~2 (which has better statistics than obs.~1),
by collapsing the counts in the X (east-west) and Y (south-north)
directions, in  a rectangular region of $222\times91$ pixels around the
SGR, excluding any point sources in the field.  Since our source
  lies very close to the PN CCD gap, we used an exposure-map corrected
  image for this analysis to correct for these CCD gaps, which also
  corrects for bad pixels. The background level, shown as a black
solid line in Figure~\ref{extent}, is the mean value of the total
counts in two regions taken at rectangular areas away from the source
in both directions. The profile is centered at the SGR central pixel,
with the dotted lines representing the  25\arcsec\ point-like source
emission, i.e., the SGR, and the dashed lines showing the extent of
the extended emission. It is clear from  both panels of
Figure~\ref{extent} that the extended  emission is asymmetrical. In
the X-direction, the emission extends up to $\sim$165\arcsec\ to the
right of the source, but only $\sim$90\arcsec\ to the  left.  In the
Y-direction the emission extends up to 125\arcsec\ below  the source
center and only up to $\sim$45\arcsec\ above it.

Finally, we detect in Obs.~1 a weak excess emission consistent with a
point source at the position of \src. Since the emission around
  \src\ shows an excess over the PSF fit starting at 18\arcsec
  (see Figure~\ref{radprof-fig}), we estimate the count rate in a
18\arcsec-radius circle centered on the source. We find a rate of
$0.0028\pm0.0006$~counts~s$^{-1}$, which represents a detection at 
the 4.6$\sigma$ level. We also detect asymmetrical emission
west--southwest of the SGR, consistent with the shape and direction of
the post-outburst asymmetrical emission discussed above. 

We summarize our spatial analysis results in Figure~\ref{imagxray}. In
the post-outburst observation (upper and middle panels), the smallest
green circle with a 25\arcsec\ radius represents the \src\ point
source emission (taking into account the PSF). The green annulus with
inner and outer radii of 25\arcsec\ and 50\arcsec, respectively
(region A hereinafter), represents the symmetrical extended emission,
most likely a dust scattering halo (see \S~\ref{sec:spec}), similar to
the one seen in the \chandra\ post-outburst observation (K+12). Beyond
$r\sim50$\arcsec\ from the center of \src, the asymmetrical extended
emission is mostly seen towards the west--southwest of the SGR (middle
panel); we approximate this region with an ellipse of major (minor) 
axis of 145\arcsec\ (95\arcsec), respectively (region B
hereinafter). Similar asymmetrical emission is seen in the
pre-outburst \xmm\ observation with some hints of weak excess emission
at the position of the SGR (lower panel). A similar extended emission
has been reported for the \chandra\ pre-outburst observations, when
the source was in quiescence  (K+12). The asymmetrical shape argues
against a dust scattering halo origin, and its small size with the
lack of any radio counterpart makes a SNR explanation questionable. A
third option is, therefore, a wind nebula powered by the magnetar. We
will discuss these possibilities in \S~\ref{sec:discuss}.

\subsection{Timing analysis}
\label{sec:timing}

For our timing analysis, which was only performed for obs.~2, we first converted the arrival times of all 2900 events within the 25\arcsec\ source photon extraction region to the arrival times at the Solar system barycenter. We then employed a $Z^2_{\rm 1}$ test \citep{buccheri83AApulse} to search for pulsed signal from the source. We detect the pulsed signal very clearly (with a $Z^2_1$ peak of about 750) at a frequency of 0.4028466(5) Hz.  Note that the measured pulse frequency of \src\ is consistent within uncertainties with the spin ephemeris reported by K+12.

\begin{figure}[!t]
\begin{center}
\includegraphics[angle=0,width=0.48\textwidth]{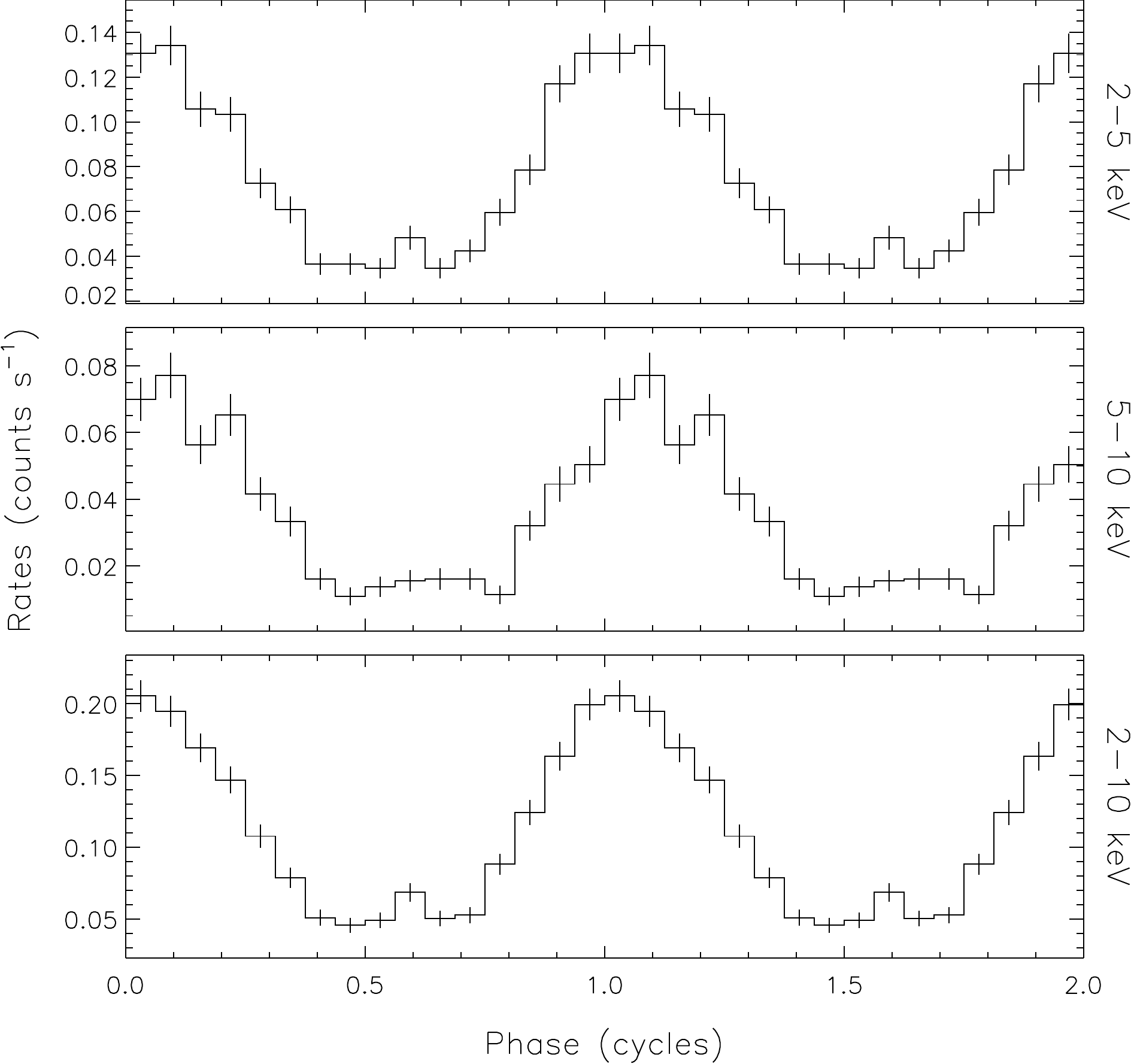}
\caption{Pulse profiles of the persistent X-ray emission of \src, accumulated between $2-5, 5-10$ and $2-10$\,keV from top to bottom.}
\label{figpulse}
\end{center}
\end{figure}

We then investigated the energy and time dependence of the pulse
profiles. Figure~\ref{figpulse} shows the background subtracted pulse
profiles in the $2-5, 5-10$ and $2-10$\,keV, respectively, from top to
bottom panels. We find  that the pulse fraction shows a hint of energy
dependence: it is (57$\pm$13)\%\ in the $2-5$ keV band and
(70$\pm$17)\%\ in $5-10$ keV. The pulsed fraction in the $2-10$ keV
band is (60$\pm$15)\%. This value is marginally lower than the value
of $85\pm10$\%\ obtained from the \chandra\ observation (K+12),
indicating a decline in pulse fraction in over about one month. We
also searched for pulse profile evolution in time by splitting the
effective duration of the \xmm\ pointing into three parts and
generating the pulse profile in each segment in the $2-10$ keV
range. We find no significant variation of pulse shape throughout the
observation as well as between the \xmm\ and \chandra\ observations.

\subsection{Spectral analysis}
\label{sec:spec}

\subsubsection{Post-outburst observation}

\begin{figure}[!t]
\begin{center}
\includegraphics[angle=0,width=0.5\textwidth]{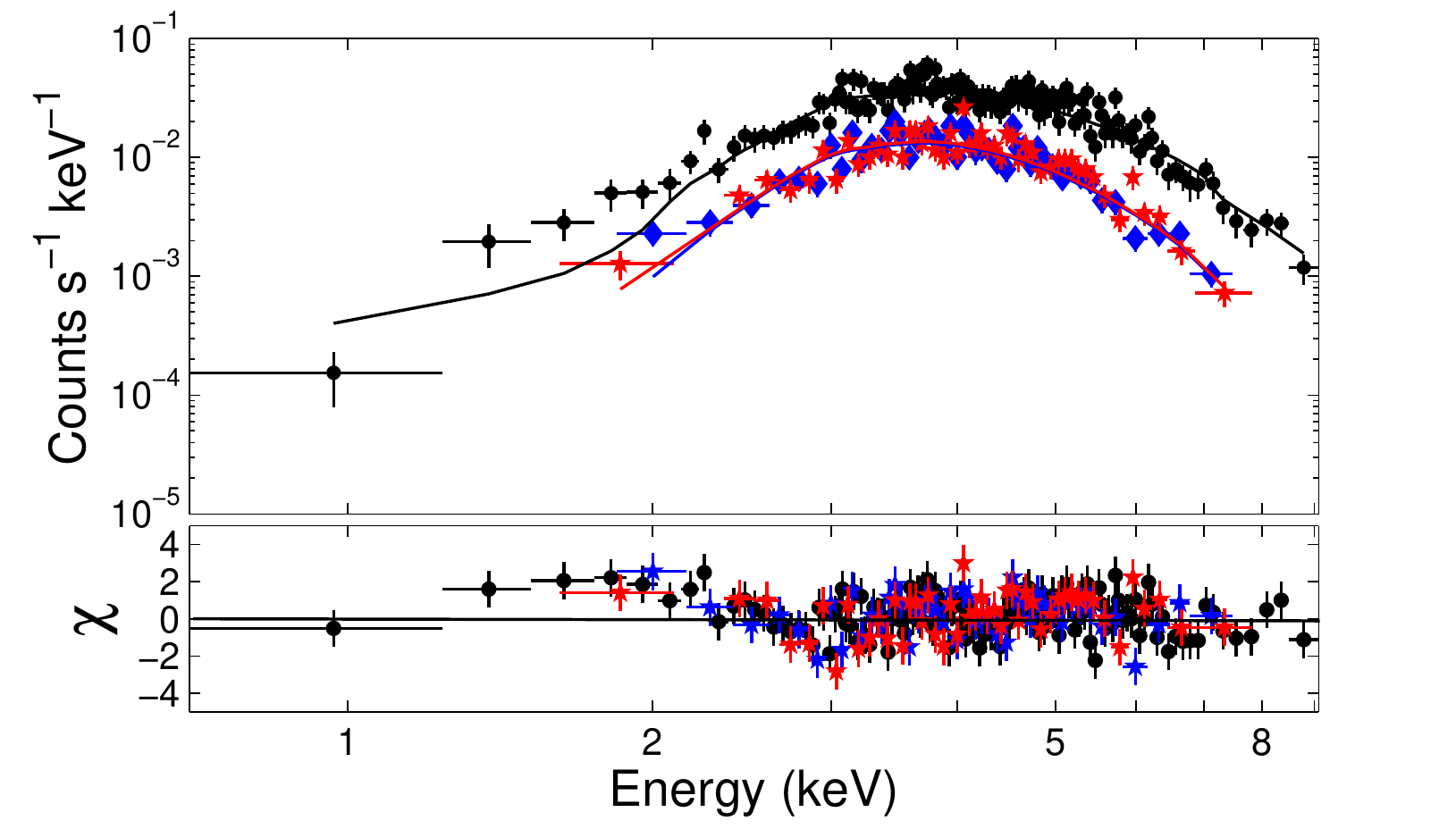}\\
\includegraphics[angle=0,width=0.5\textwidth]{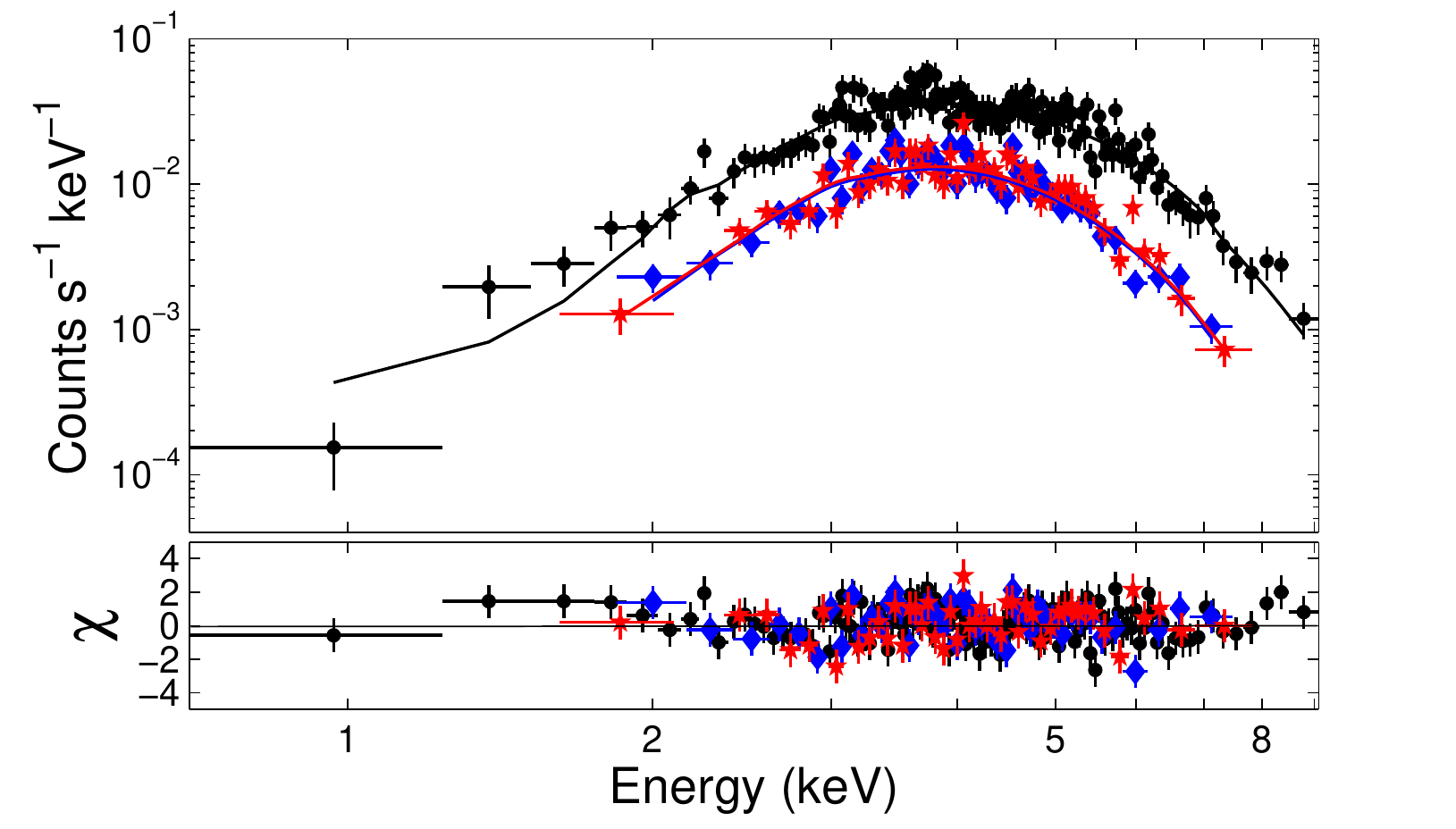}
\caption{{\sl Upper panel.} Data and power-law fit to the \src\
  post-outburst \xmm\ data. {\sl Lower panel.} Data and blackbody fit to the \src\
  post-outburst \xmm\ data. In both panels, black dots, blue diamonds, and red stars represent PN, MOS1, and MOS2 data, respectively. Residuals are shown in terms of sigma.}
\label{swiftspec-fig1}
\end{center}
\end{figure}

\begin{figure}[!t]
\begin{center}
\includegraphics[angle=0,width=0.5\textwidth]{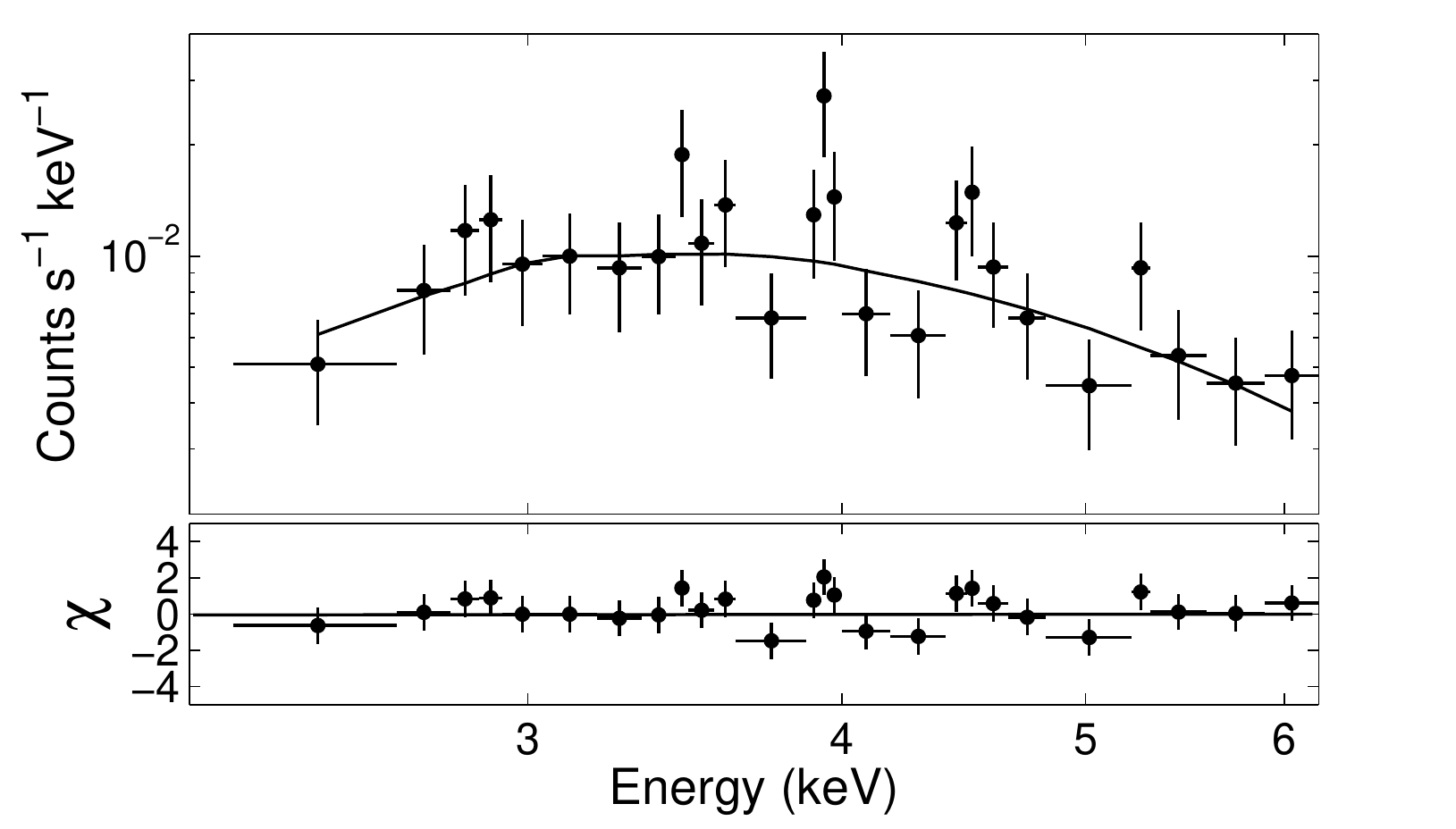}\\
\includegraphics[angle=0,width=0.5\textwidth]{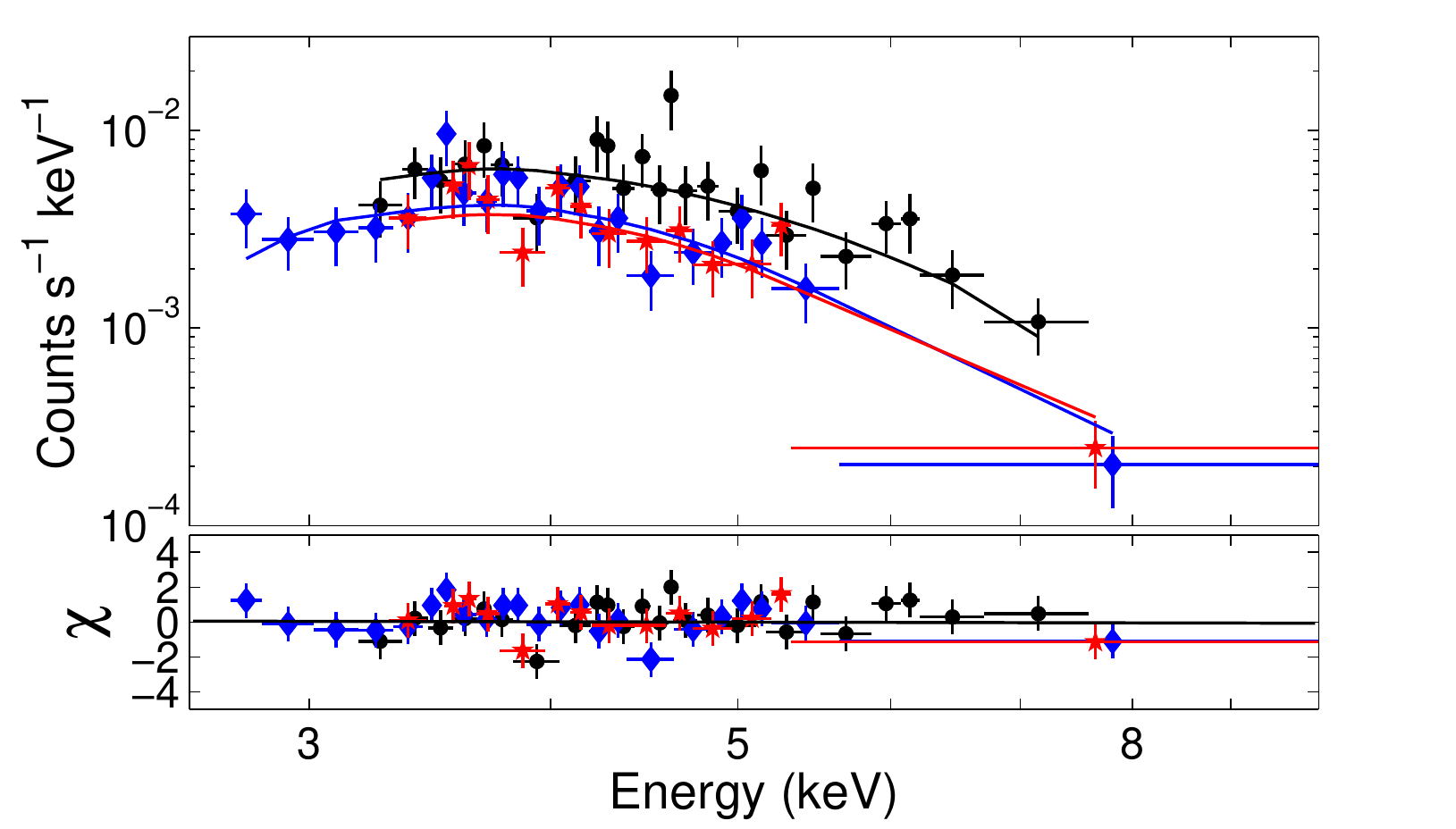}
\caption{{\sl Upper panel.} Data and power-law fit to region B
  during the post-outburst \xmm\ observation. {\sl Lower panel.} Data and power-law fit to region A
  during the post-outburst \xmm\ observation. Black dots, blue diamonds, and red stars represent PN, MOS1, and MOS2 data, respectively. Residuals are shown in terms of sigma.} 
\label{swiftspec-halo}
\end{center}
\end{figure}

\begin{figure}[!t]
\begin{center}
\includegraphics[angle=0,width=0.5\textwidth]{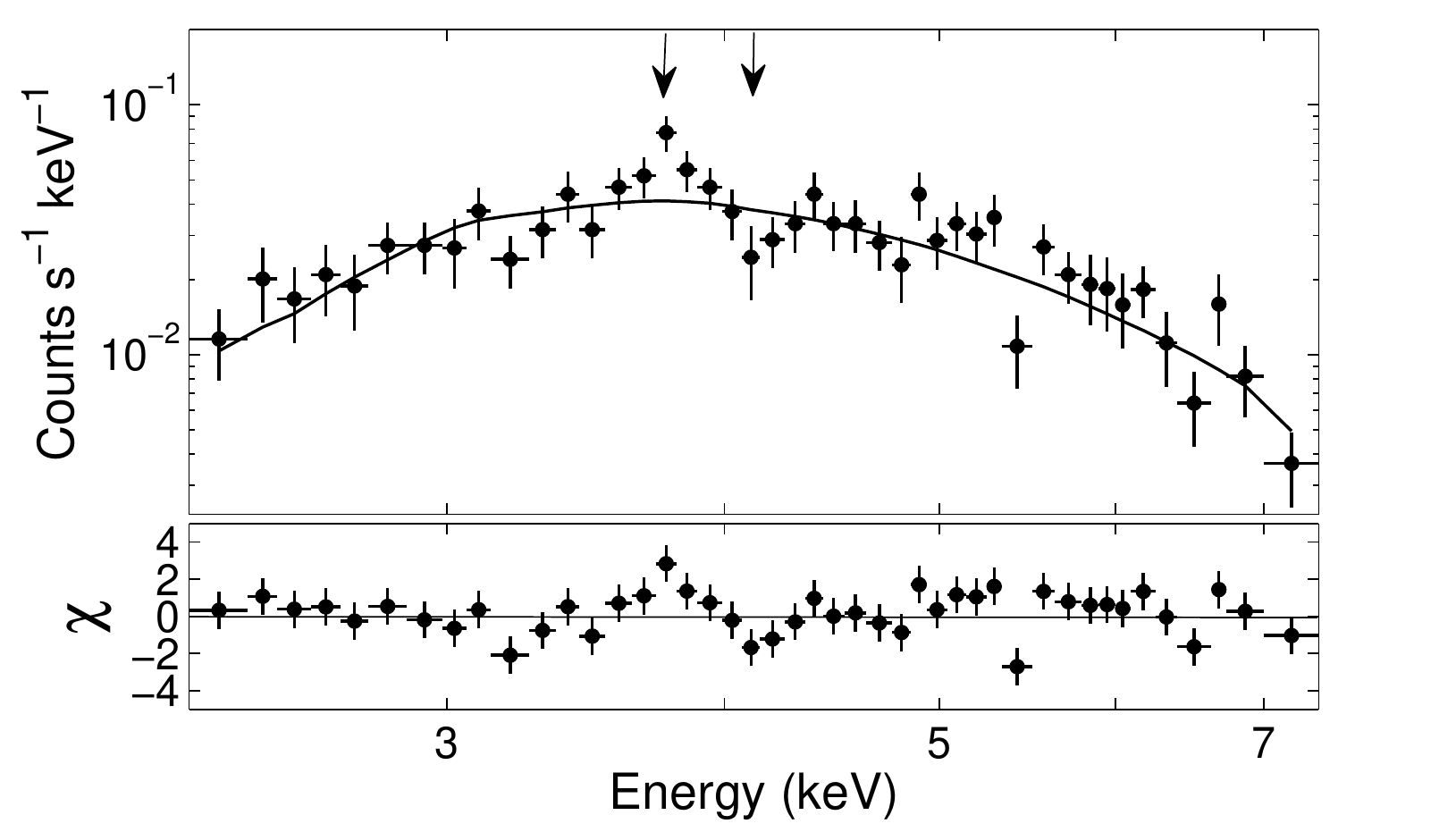}
\caption{Data and PL fit to the second time-segment (see text) of the post-outburst \xmm\ observation of \src. Residuals are shown in
  terms of sigma. The arrows indicate potential emission and absorption features at $\sim$3.7 and 4.2 keV, respectively.}
\label{seg2-spectrum}
\end{center}
\end{figure}

Based on  our radial profile analysis of obs.~2, we extracted  the
spectra of \src\  in a circular region with a radius of 25\arcsec\
from the PN camera and with a radius of 20\arcsec\  from the MOS1/MOS2
cameras (extended emission started at 20\arcsec\ from the center of
the SGR in the MOS cameras), collecting 2900 and 1020 counts,
respectively. Background  events were extracted from source--free
circles with the  same radii as for the source and on the same CCD,
resulting in 56 and 32, PN and MOS1/MOS2 background counts
respectively. The spectra were then grouped to have a minimum of 25
counts per bin. Finally, we  made sure that the point source spectrum
was  not  affected by pile-up using the \xmm\ SAS task {\sl
  epatplot}. Table~\ref{swiftspec-tab} includes the results of our
spectral analysis of the point source and both extended regions (see
below).

We fit the point source (\src) spectrum with an absorbed power-law (PL) and with an absorbed blackbody (BB) model. The latter gave a better fit, with a reduced $\chi^2$ of 1.04 for 232 d.o.f., corresponding to  an improvement of  26 in $\chi^2$ for the same number of d.o.f. From the BB fit we estimate the emitting area radius to be $R=(0.24\pm0.02)d_4$~km, where $d_4=d/4$~kpc, consistent with the value derived from the \chandra\ data taken $\sim$1~month earlier. 
Table~\ref{swiftspec-tab} gives the PL and BB best-fit parameters, and the absorbed fluxes and luminosities. Figure~\ref{swiftspec-fig1} upper (lower) panel shows the best-fit PL (BB) model and the residuals in terms of sigma. In each panel of Figure~\ref{swiftspec-fig1} the upper (black dots) fits are the EPIC-PN data and the two lower fits (blue diamonds and red stars) are the MOS1 and MOS2 data. We note here that the  fluxes and luminosities of \src, are half the values derived from the \chandra\ data almost a month earlier (K+12). Finally, we note that a more complex, two-component model, typically used to fit magnetar X-ray spectra, is not required by the data.

\begin{table*}[t]
\caption{Spectral model parameters, fluxes and luminosities of \src\ and its surrounding medium.}
\label{swiftspec-tab}
\newcommand\T{\rule{0pt}{2.6ex}}
\newcommand\B{\rule[-1.2ex]{0pt}{0pt}}
\begin{center}{
\resizebox{0.95\textwidth}{!}{
\begin{tabular}{l c c c c c c c c}
\hline
\hline
Source & Model \T\B & N$_{\rm H}$ & $\Gamma$ & $kT$ & $N^a$ or $R^b$ & $\chi_{\nu}^2$/d.o.f. & $F_{\rm 2-10\,keV}$ absorbed & $L_{\rm 2-10~keV}^c$ \\
   \T\B &   & ($10^{22}$~cm$^{-2}$) & & (keV) &    & & ($10^{-12}$~erg~cm$^{-2}$~s$^{-1}$) & ($10^{34}$~erg~s$^{-1}$) \\
\hline
\src\ (post-outburst)    \T  & PL &  $24\pm1$      & $4.2\pm0.1$&$\ldots$ & $5.67_{-0.01}^{+0.02}$ &  1.15/232  &  $1.25^{+0.02}_{-0.03}$ &$1.6_{-0.1}^{+0.2}$ \\
\src\ (post-outburst)    \T  & BB &  $12.9\pm0.6$ & $\ldots$           &  $0.96\pm0.02$ & $0.24\pm0.02$& 1.04/232  &  $1.19^{+0.03}_{-0.04}$ &$0.16\pm0.01$ \\
\src\ (pre-outburst)$^d$        \T \B & PL   & $24 (fixed)$    & $4.2 (fixed)$ & $\dots$ &$\dots$ &$\dots$&  $0.04$ & $0.07$ \\
\hline
Region A (post-outburst) \T & PL  &  $25_{-5}^{+6}$ & $4.5_{-0.6}^{+0.7}$  & $\ldots$ & $1.48\pm0.02$ & 0.9/57 &  $0.19\pm0.02$ &$0.3_{-0.2}^{+0.5}$ \\
Region A (post-outburst)$^e$ \T & PL  &  $31_{-9}^{+10}$ & $5.0_{-0.9}^{+1.0}$  & $\ldots$ & $3.20^{+0.02}_{-0.01}$ & 0.9/57 &  $0.16\pm0.02$ &$0.4_{-0.2}^{+0.5}$ \\
Region A (pre-outburst) \T \B & PL & $13_{-6}^{+8}$ &$1.7_{-1.1}^{+1.4}$&$\ldots$ & $0.005^{+0.007}_{-0.003}$ &1.3/8& $0.12_{-0.05}^{+0.06}$ & $0.04_{-0.01}^{+0.02}$ \\  
\hline
Region B (post-outburst) \T & PL  &   $15\pm5$     & $3.4_{-0.9}^{+1.0}$       & $\dots$ & $0.3\pm0.1$ &1.0/23   &  $0.35\pm0.06$ & $0.21_{-0.06}^{+0.15}$ \\
Region B (post-outburst)$^f$ \T & PL  &   $17\pm4$     & $3.2_{-0.6}^{+0.7}$       & $\dots$ & $0.2\pm0.1$ &0.9/46   &  $0.35\pm0.04$ & $0.21_{-0.06}^{+0.10}$ \\
Region B (pre-outburst)   \T \B & PL  & $16 (fixed)$     & $3.5\pm0.6$       & $\dots$& $0.2^{+0.2}_{-0.1}$& 1.7/19 &  $0.15_{-0.05}^{+0.06}$  & $0.10_{-0.03}^{+0.04}$ \\
\hline
\end{tabular}}}
\end{center}
\begin{list}{}{}
\item[{\bf Notes.}]$^{(a)}$ PL normalization in units of 10$^{-2}$
  photons cm$^{-2}$ s$^{-1}$ keV$^{-1}$ at 1 keV.\\ $^{(b)}$ BB radius,
  in units of km. \\$^{(c)}$ $2-10$\,keV power-law luminosity
  or bolometric BB luminosity ($\pi R^2 \sigma T^4$), assuming a
  source distance of 4~kpc \citep{tian07ApJ:1834}.\\ $^{(d)}$ Fluxes
  and luminosities converted from the count rate in
  Section~\ref{spa-ana}  using {\sl PIMMS}, assuming the corresponding
  spectral parameters.\\ $^{(e)}$ Spectral results including the
  possible contribution from region B (see Section~\ref{sec:spec}).\\
  $^{(f)}$ Spectral parameters derived using a modeled background as
  described in Section~\ref{sec:obs}.
\end{list}
\end{table*}

We then binned the spectra of the point source to the PN spectral
resolution and searched for potential line-like features in the
time-integrated and the time-resolved spectra. The time-integrated
spectrum revealed two possible lines (absorption and emission) between
$3-5$\,keV. To investigate the lines, we first added a Gaussian
emission profile with a best-fit energy of 3.7~keV, which reduced
$\chi^2$ by 8, for 3 d.o.f. The addition of an absorption line with a
best-fit energy of 4.2~keV resulted in an equal improvement. Adding
both lines together does not improve the spectral fit further. We then
performed Monte Carlo Simulations (MCS) to rigorously assess the
significance of these spectral features. We took the best-fit absorbed
PL model as our null hypothesis. We simulated 1000 spectra based on
this model with the XSPEC {\sl fakeit} command, and fitted each
spectrum with the null hypothesis model. We then added an absorption
line to the model ({\sl gabs} in XSPEC) and re-fit the spectrum. For
each simulated spectrum, we recorded the $\Delta\chi^2$ between the
null hypothesis PL model and the PL + absorption feature model, and
compared the values to the real $\Delta\chi^2$. This procedure
resulted in an absorption line significance at only the 90\%\
confidence level. Including an emission line at 3.7\,keV, instead of
an absorption line, gave the same level of significance. We note that
this significance level is insensitive to the null hypothesis model
since an absorbed BB gave similar results (95\% confidence level). We
conclude that the lines are not significant in the time-integrated
spectrum of \src.

Next, we performed both time-resolved and phase-resolved spectroscopy
to investigate whether there are specific intervals (phases) where the
lines are more prevalent. For the former case, we split the $\sim$24\,ks
observation into four equal segments and fit each of the four spectra
with an absorbed PL model. We find that the source spectrum is
constant throughout the observation. Only in segment two
($6.75-13.50$\,ks) we see evidence for the presence of an emission
line at 3.8\,keV (Figure~\ref{seg2-spectrum}, first arrow). A MCS showed
that the line is significant at the 98.5\%\ confidence level. A MCS of
an absorbed BB spectrum with the same emission line resulted in a
$\sim$99\%\ confidence level. The significance is too low to claim
a firm line detection; more sensitive observations during a new source
burst active episode could provide better statistics.

To perform phase-resolved spectroscopy, we rebinned by a factor of two the profiles of Figure~\ref{figpulse}, starting at phase=0, which resulted in a total of 8 bins. We then fit each spectrum with an absorbed PL (with $N_{\rm H}$ fixed to the best-fit value, see Table~\ref{swiftspec-tab}). We find no variations across the spectra within uncertainties.

The high hydrogen column density that we derive for the source
suggests that there should be an accompanying dust scattering halo
emission \citep{predehl95aa:rosat}. Such emission must be symmetrical
except for a very unusual dust distribution. Hence, we extracted a
spectrum from an annular region $25\arcsec\le r\le50\arcsec$
(region A), from PN, MOS 1 and MOS 2. The source contribution to the
extended emission is supposed to be minimal, including at most 20\%\
from the outer wings of the EPIC-PN
PSF\footnote{http://xmm.esac.esa.int/external/xmm\_user\_support/documentation/uhb\_2.1/}. Region
A, on the other hand, could contain some
  contribution from the more extended asymmetrical emission
  (see Section~\ref{spa-ana}). Hence, we modeled the region A
  spectrum, first, as a separate component, and second taking into
  account the possible contribution from region B
  (see below). We find that the spectrum of Region A is well fit with
  an absorbed PL in both cases (Figure~\ref{swiftspec-halo}), with
  similar $N_{\rm H}$ and photon indices. These parameters are also
  consistent with those of the SGR within the uncertainties. These
  results are presented in Table~\ref{swiftspec-tab} and discussed in
  \S~\ref{sec:discuss}.

We then extracted the PN spectrum of the asymmetrically extended
emission (hereafter region B) using an ellipse with a semi-major/minor
axis of 145\arcsec\ and 95\arcsec, respectively, which encloses the
elliptical region shown in Figure~3. We excluded the \src\ and
region~A extraction areas. The $0.5-10$~keV spectrum is adequately fit
with an absorbed PL (Figure~\ref{swiftspec-halo}) with a hydrogen
column density $N_{\rm H}$ and photon index $\Gamma$ consistent within
uncertainties with those of the point source and region A
spectra. Fixing $N_{\rm H}$ to the best fit value better constrains 
$\Gamma=3.4_{-0.3}^{+0.2}$; this value is smaller than the point
source index at the $3\sigma$ level. All fit parameters and absorbed
fluxes and luminosities are given in Table~1.


\subsubsection{Pre-outburst observation}

The 2005 \xmm\ observation of the field of \src\ shows a weak
point-like source at the position of \src. We collected 45 counts from
the 18\arcsec-radius circle around \src\ as shown in the lower panel
of Figure~\ref{imagxray}, not enough for a proper spectral
analysis. We, therefore, assumed the same spectral parameters as in
the post-outburst observation, to derive the $2-10$~keV absorbed flux
and luminosity listed in Table~1. A photon index $\Gamma=3.0$, assuming
the source X-ray spectrum hardens with declining flux \citep[e.g.,
][]{gogus10ApJ:1833}, would only decrease the luminosity by a factor
of 1.5.

Next, we collected $\sim$100 counts from region A and binned the
spectrum at 15~counts/bin. We then fit it with an absorbed PL and found that the
absorbing column and $\Gamma$ are consistent, within uncertainties,
with the post-outburst values for this region (see also Table 1).

We also extracted the $0.5-10$ keV spectrum of region B using the same
elliptical region as above (Figure~\ref{imagxray}), excluding a
50\arcsec-radius circle around the \src\ position. This resulted in a
total of $\sim90$~counts. Because of the low statistics we grouped the
spectrum to have 40 counts per bin, achieving a S/N ratio of $\sim2$
(this low S/N is due to the large background of MOS1/MOS2 compared to
the extended emission photon flux). We fit the spectrum with an
absorbed PL. We also fixed the column density to the best-fit value,
$N_{\rm H}=1.6\times10^{23}$~cm$^{-2}$, and found $\Gamma=3.5\pm0.6$,
consistent with the post-outburst extended emission value. The
absorbed flux and luminosity of Region B are roughly a factor of two
lower than their post-outburst values. These results are also
discussed in Section~\ref{sec:discuss}.

\section{Discussion}
\label{sec:discuss}

\subsection{\src}

The effects of bursting activity on the magnetar persistent X-ray flux
have been discussed by several authors. The increase of the source
intensity during bursting episodes is also often accompanied by
spectral variability \citep[e.g. ][]{vasisht00ApJ:ax1845,
  gotthelf04ApJ:xte1810, gogus10ApJ:0501}. It would then be reasonable
to assume that the detection of \src\ in the 2005 \xmm\ observation at
$F_{\rm 2-10\,keV}\approx10^{-13}$~erg~cm$^{-2}$~s$^{-1}$, could be
due to a bursting episode that had occurred prior and close to that
observation (if such an episode comprised only one burst similar to
the 2011 episode, it could have easily been missed by {\it Swift},
which was the only all sky monitor in the $25-350$\,keV range at the
time). Indeed, assuming a (constant) flux decay trend between 2005 and
2009 similar to the one exhibited by the source after its 2011
outburst ($\alpha=-0.5$, Figure~\ref{nebulavar}) results in an
expected flux level in 2009, consistent with the estimated upper limit
of $10^{-15}$~erg~cm$^{-2}$~s$^{-1}$ (K+12).

However, there maybe other sources of neutron star surface heating
that might not result in SGR bursts, such as was the case of the
transient magnetar SGR~J$1810-197$ \citep{ibrahim04ApJ:xtej1810}. The
source was serendipitously discovered with \rxte\ as a transient
during observations of a nearby magnetar (SGR~J$1806-20$); the
increase of its X-ray flux was not associated with any bursting
activity during that period. This behavior could be explained within
the framework of the twisted magnetosphere model of
\citet{thompson02ApJ:magnetars} as follows. Variations of the twist
angle of the magnetic field lines would lead to a sudden release of
energy accompanied by possible changes in the cyclotron resonant
scattering depth in the magnetosphere and heating of the neutron star
surface. Heating by such a $B-$field reconfiguration should also be
associated with sharp spectral changes. Unfortunately, with the
currently available data we cannot distinguish between the two
scenarios.

\begin{figure}[!th]
\begin{center}
\includegraphics[angle=0,width=0.48\textwidth]{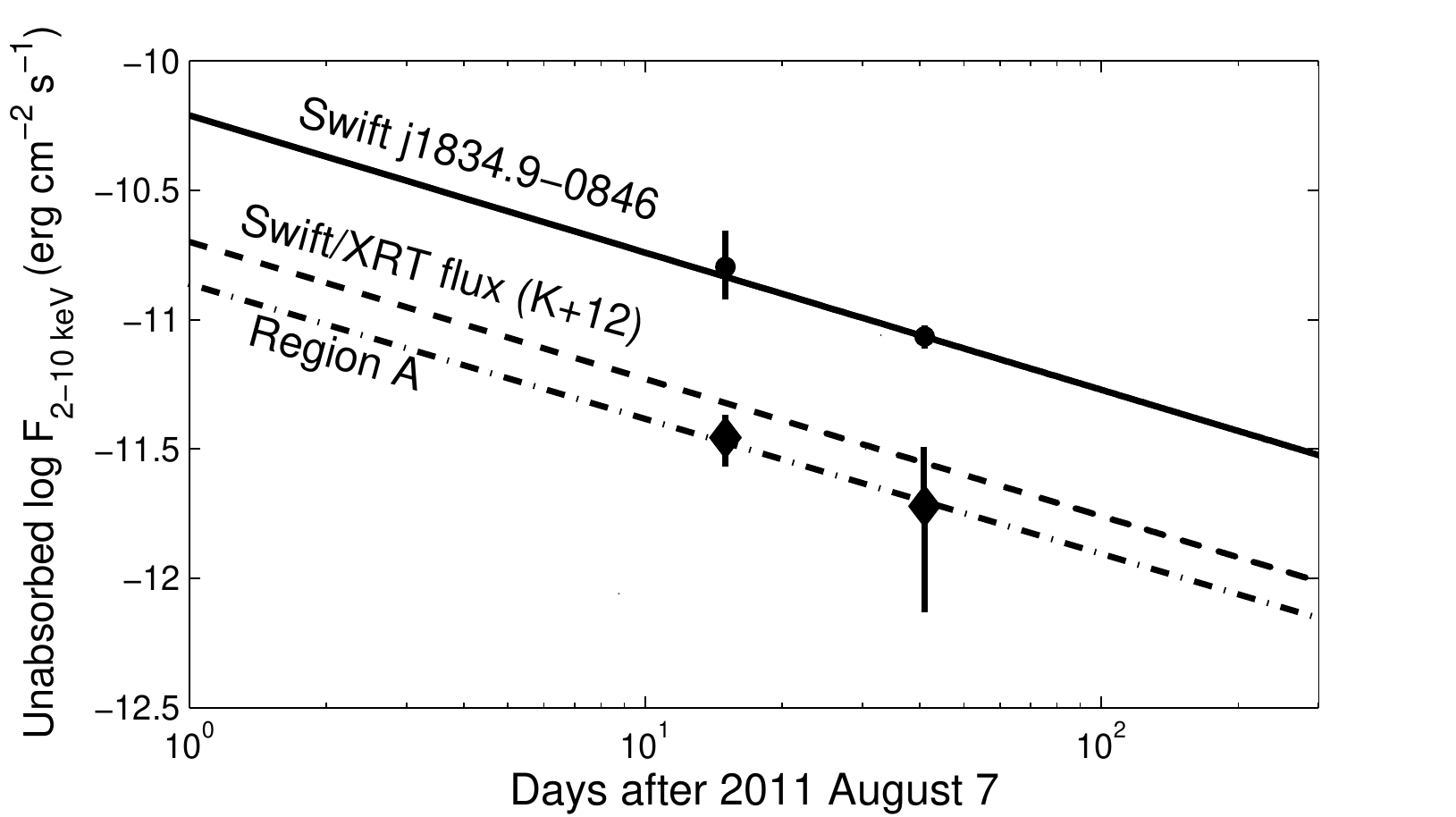}
\caption{The post-outburst persistent X-ray light curve of \src\ based
  on 48 days of \swift/XRT data (dashed-line, K+12); day 1 corresponds
  to the \swift\ trigger. The dots represent the \chandra\ and \xmm\
  post-outburst point source fluxes ($2-10$\,keV), respectively, while
  the diamonds represent the fluxes of Region A during the same
  observations. The dashed line represents the \swift/XRT decay slope of
  $-0.5$; the solid and dot-dashed lines are decay trends of the point
  source and Region A with the same slope.}
\label{halosrcvar}
\end{center}
\end{figure}

Magnetar X-ray spectra are usually fit by a two component model, e.g.,
two BBs with temperatures $kT_{1}\sim0.3$~keV and $kT_{2}\sim0.8$~keV,
or a BB and a PL with $kT\sim0.5$~keV and $\Gamma\sim3.0-4.0$
\citep[e.g., ][]{mereghetti05ApJ:sgr1806,
  halpern05ApJ:xte1810,tiengo08ApJ:cxouj010043, bernardini09:axp1810,
  bernardini11:1E1547, rea09MNRAS:sgr0501,
  gogus11ApJ:1900,woods07ApJ:1806, kouveliotou03ApJ:1627,
  kouveliotou01ApJ:1900}. The 2005 pre-outburst spectral properties of
the source could not be inferred due to very low statistics. The
post-outburst X-ray spectrum of \src\ seems unusual at first glance, as
it is well fit by a single, heavily absorbed ($N_{\rm
  H}\sim10^{23}$~cm$^{-2}$) component, either a blackbody with
$kT=1.1$~keV or a power-law with $\Gamma=4.2$ (see also K+12). It
could be that we see here the effects of the environment within which
\src\ resides; e.g., dense giant molecular clouds (GMCs,
\citealt{tian07ApJ:1834}), which, in principle, could absorb the soft
part of the spectrum, eliminating the requirement of a soft spectral
component \citep[see also ][]{esposito11MNRAS:1833}.

The single BB spectral model for \src\ gives a small decrease in the
BB temperature ($\Delta kT=0.14\pm0.06$~keV), and a consistent BB
emitting area radius ($\Delta R=0.02\pm0.05$) between the \chandra\
and \xmm\ post-outburst observations separated by a month, similar to
the behavior of XTE~J$1810-197$ \citep{woods05ApJ:xte1810}. The BB
fluxes between the two observations are consistent with the same
power-law decay $\alpha\approx-0.5$, estimated using the PL fits. K+12
discussed the possibility of a hot spot emitting thermal radiation at
the surface of the neutron star, noting that in such a scenario it
would be difficult to explain the high pulsed fraction due to light
bending in the neutron star gravitational field, unless the radiation
is anisotropic, having a narrow peak along the magnetic field
direction \citep{pavlov94AA:magnetars}. 

\subsection{A Halo around \src: Region A}

The spectrum and flux of the symmetrical extended emission (region A)
fits well a dust scattering halo interpretation. First, the heavy
absorption ($N_{\rm H}\approx10^{23}$~cm$^{-2}$) towards the source,
inferred from the X-ray spectral fits, should cause the scattering
of the point source X-ray emission, resulting in a dust scattering
halo. Since the scattering cross section of the dust particles is
proportional to $E^{-2}$, a halo is expected to have a softer spectrum
than the illuminating source, i.e., \src. Indeed, in Obs.~2, the spectrum
of region A is marginally softer than the \src\ spectrum (although
consistent within the uncertainties, see Section 3.3 and Table~1).
Second, a dust scattering halo is expected to vary in flux
proportionally to the illuminating source flux
\citep{mathis91ApJ:halo}, with a time lag depending
on the distance of the scattering material from the source
\citep{mauche86ApJ:halo,olausen11ApJ:sgr1550}. This trend is evident
from Figure~\ref{halosrcvar}, which shows the flux evolution of
region~A and \src, between the post-outburst \chandra\ (K+12) and
\xmm\ observations (diamonds). Finally, we estimate the
  fractional intensity of the halo during Obs.~2 to be $I_{\rm
    frac}=F_{\rm halo}/(F_{\rm halo}+F_{\rm source})=0.20_{-0.10}^{+0.25}$.

During Obs.~1 the spectrum of region A was harder,
  $\Gamma=1.7^{+1.4}_{-1.1}$, with a fractional intensity $I_{\rm
    frac}=0.36^{+0.2}_{-0.1}$, somewhat higher than, but consistent within
  the error bars with the $I_{\rm frac}$ calculated for Obs.~2.
However, the \src\ spectrum during obs.~1 is unknown due to the poor
statistics. The harder spectrum during obs.~1 could then be explained
if there were another component contributing to the flux in region
A. Indeed, the flux of region B (the putative MWN, see Section 4.3)
dominates the emission from the vicinity of \src\ during Obs.~1
(Table~1), which could explain both the hard spectrum and the slightly
higher $I_{\rm frac}$ seen during this observation. Another
explanation could be that the \src\ spectrum during Obs.~1 is much
harder than it is during obs.~2, which would make the region A
spectral shape consistent with a solely dust scattering halo
explanation.

\subsection{Asymmetrical Extended emission (Region B): a MWN?}

\begin{figure}[!th]
\begin{center}
\includegraphics[angle=0,width=0.48\textwidth]{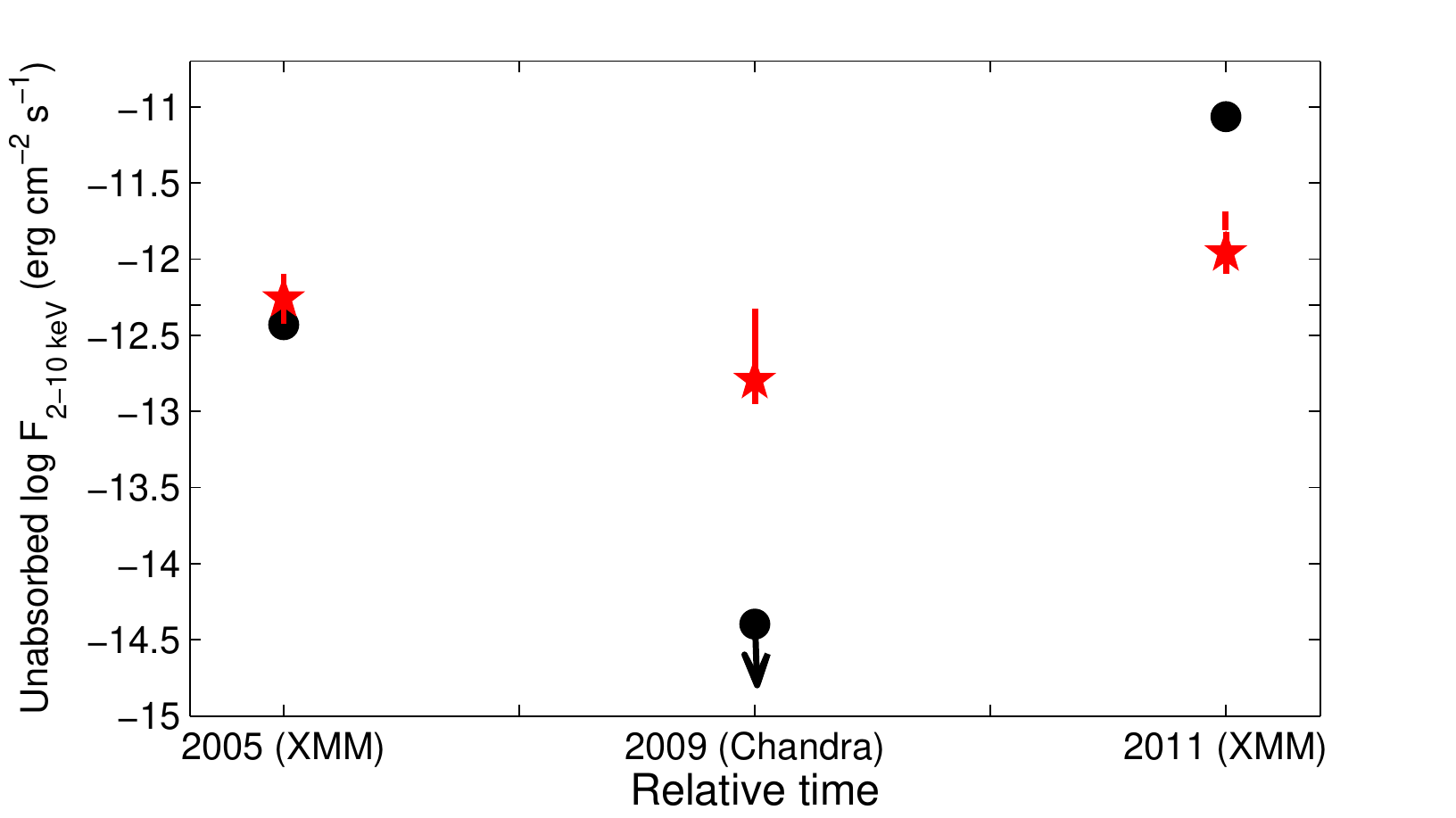}
\caption{Long term light curves of the fluxes ($2-10$\,keV) of \src\ (black dots) and Region B (red stars).}
\label{nebulavar}
\end{center}
\end{figure}

Rotation Powered Pulsars (RPP) with magnetic fields
$B\sim10^{11-13}$~G and periods $P\lesssim1$~s are believed to lose
their rotational energy in the form of a relativistic magnetized particle
wind. Pulsar wind nebulae (PWNe) are often observed around these
pulsars and are believed to be the synchrotron radiation of the
shocked wind \citep[see ][for
reviews]{kaspi06csxs:PWN,gaensler06ARA:PWN, kargaltsev08PWN}. The
efficiency at which the rotational energy loss of a pulsar,
$\dot{E}_{\rm rot}$, is radiated by the PWN is characterized by
$\eta_{\rm X}=L_{\rm X, PWN}/\dot{E}_{\rm rot}$, which ranges from
10$^{-6}$ to 10$^{-2}$. Magnetars, on the other hand, have longer spin
periods and lower $\dot{E}_{\rm rot}$ values, making the production of
a steady and bright rotationally-powered nebula unlikely. Nonetheless,
\citet{thompson98PhRvD:mag} showed that particle outflows, either
steady or released in short periods of time due to the flares, could
be driven by Alfv\'en waves \citep[see also
][]{harding99ApJ:mag}. Furthermore, a jetted baryonic outflow was
observed in the radio wavelengths after the GF of SGR~J$1806-20$
\citep{gaensler05Natur:1806,fender06MNRAS:1806}. These processes could
lead to the emergence of nebulae around magnetars.


There has not been yet a ubiquitous detection of a magnetar wind
nebula (MWN) in X-rays, but ``magnetically powered'' nebulae around
pulsars with relatively high magnetic fields have been
suggested. \citet{rea09ApJpwn} reported that the nebula around the
rotating radio transient RRAT\,J$1819-1458$ has a nominal X-ray
efficiency $\eta_{\rm X}\approx0.2$, too high to be rotationally
powered. The authors suggested that the occurrence of the nebula might
be connected with the high magnetic field ($B=5\times10^{13}$~G) of
the pulsar.

\begin{figure}[t]
\begin{center}
\includegraphics[angle=0,width=0.49\textwidth]{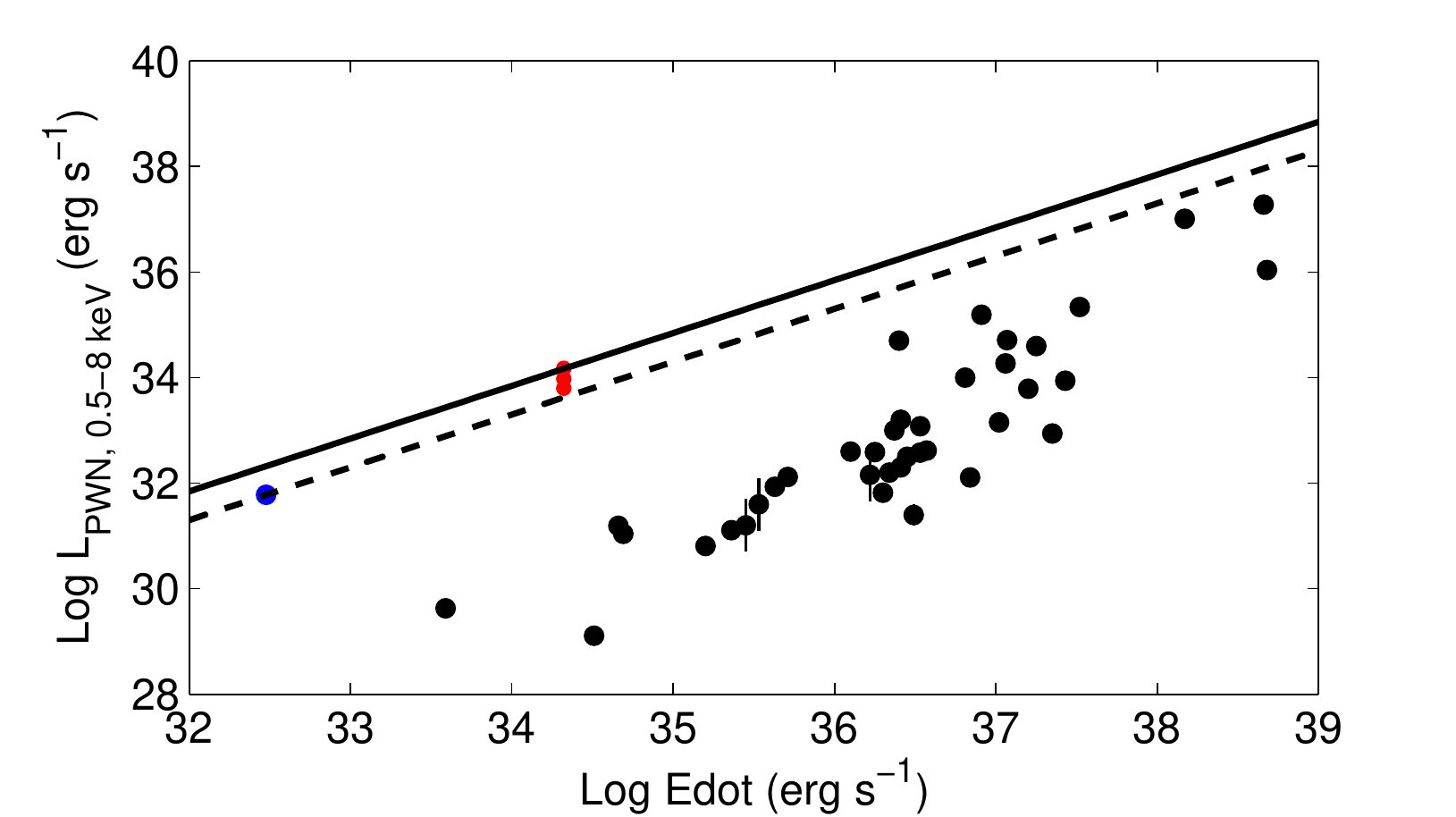}
\caption{Luminosity of normal PWNe as a function of the rotational
  energy loss of their corresponding pulsars. Data presented as black
  dots are taken from \citet{kargaltsev08PWN}, whereas the blue star
  represents the high B source RRAT\,J$1819-1458$
  \citep{rea09ApJpwn}. The dashed line represents the $\eta_{\rm X}=0.2$ of
  RRAT\,J$1819-1458$, and the solid line represents the $\eta_{\rm X}=0.7$ of
  \src. The three red dots represent the luminosity of the candidate
  MWN around \src\ at the detected epochs. [Figure adapted from
  \citet{rea09ApJpwn}].}
\label{LpwnEdot}
\end{center}
\end{figure}

The nebula around \src\ shares some characteristics with the nebula
around RRAT\,J$1819-1458$. The X-ray efficiency of the \src\ nebula is
very high, $\eta_{\rm X}\approx0.7$, for a $0.5-8$\,keV luminosity of
$1.5\times10^{34}$~erg~s$^{-1}$\footnote{We have chosen the 0.5-8 keV
  energy range to enable comparison with the efficiency of
  RRAT\,J$1819-1458$ and other pulsars,
  Figure~\ref{LpwnEdot}.}. Considering the source's relatively low
rotational energy loss ($\dot{E}_{\rm   rot}=2.1\times10^{34}$
erg~s$^{-1}$), it is in the low-$\dot{E}_{\rm   rot}$/high-$L_{\rm X,
  PWN}$ region in Figure~\ref{LpwnEdot}, similar to
RRAT~J$1819-1458$. Moreover, the nebula around \src\ shows small flux
variability (owing to large uncertainties) between the three different
epochs (Figure~\ref{nebulavar}). Its flux slightly decreased, although
within uncertainties, when the source went to quiescence in 2009 ($F_{\rm
  X}<10^{-15}$~erg~s$^{-1}$), then increased by a factor of 7 (at the
$\sim$2 sigma level) after the September 2011 outburst, in line with a
variable wind nebula scenario.


An obvious difference between the MWN around \src\ and the ``usual''
PWNe is the very soft spectrum of the former, $\Gamma =3.5\pm0.6$,
compared to $\Gamma\sim1 - 2$ of PWNe of RPPs. It is worth pointing
out that the nebula around RRAT~J$1819-1458$ also shows a soft
spectrum, $\Gamma=3.0\pm 0.5$, which suggests that the two nebulae are
in some respects similar, in particular, the electrons are accelerated
by similar mechanisms (we note, however, that the nebula around
  RRAT~J$1819-1458$ is about 10 times smaller in size than the nebula
  around \src, for similar distances). For the most plausible
assumption that we are observing synchrotron radiation of relativistic
electrons, this large index implies a very steep electron spectrum,
with a slope $p = 2\Gamma - 1 \approx 6$. What could produce such an
electron population? A different mechanism (other than the typically
invoked Fermi mechanism) of electron acceleration, such as, e.g.,
magnetic field line reconnection might be at work. We can only
conjecture that the twisted magnetic field model by
\citet{thompson02ApJ:magnetars} could lead to reconnection,
facilitating the production of the required electron population
distribution.

We can estimate the termination shock radius $R_s$ depending on our
assumptions about the energy flux provided by the magnetar. In
quiescence, the balance of pressures $\dot{E}_{\rm rot}/(4\pi f c
R_s^2) = p$, where $4\pi f$ is the solid angle in which the wind
(including the Poynting flux) is blowing ($f =1$ for an isotropic
wind), and $p$ is the ambient pressure (this equation assumes that the
magnetar's speed is essentially subsonic). For the $\dot{E}_{\rm rot}
= 2.1\times 10^{34}$ erg s$^{-1}$, this equation gives $R_s =
2.4\times 10^{16} f^{-1/2} p_{-10}^{-1/2}$ cm, where $p_{-10}$ is the
pressure in units of $10^{-10}$ erg cm$^{-3}$. This corresponds to the
angular size of $0\farcs4 f^{-1/2} p_{-10}^{-1/2} d_{4}^{-1}$. Such a
small size cannot be resolved by \xmm, and it is hidden within the
dust scattering halo (Region A), assuming reasonable values for the
ambient pressure. The size of an X-ray PWN is typically a factor of a
few times larger than $R_s$ (e.g., \citealt{kargaltsev08PWN}), which
is still much smaller than the observed size of
$\sim150$\arcsec. Therefore, not only the unrealistically high
``efficiency'' $\eta_{\rm X}\sim 0.7$, but also the large size support
the hypothesis that the observed asymmetrical nebula (Region B) could
not be produced by the magnetar in quiescence via rotation-powered
wind.

When a magnetar is in an active state, the pressure of its wind
(ejected particles and magnetic fields) is much higher than that in
quiescence. In this state, the energy loss rate, $\dot{E}_{\rm
  burst}$, can be much higher than $\dot{E}_{\rm rot}$. It can be
crudely estimated as a ratio of the magnetar's X-ray luminosity in the
bursting state, $L_X = 10^{34} L_{X,34}$ erg s$^{-1}$, to some
reasonable magnetar X-ray efficiency $\eta_{\rm X} = 10^{-4}\eta_{\rm
  X,-4}$: $\dot{E}_{\rm burst} = 10^{38} L_{X,34} \eta^{-1}_{\rm X,-4}$ erg
s$^{-1}$. Using $\dot{E}_{\rm burst}$ instead of $\dot{E}_{\rm rot}$,
we obtain $R_s = 1.6\times 10^{18} L_{X,34}^{1/2} \eta_{\rm
  X,-4}^{-1/2} f^{-1/2} p_{-10}^{-1/2}$ cm, which corresponds to the
angular shock radius of $\sim$25\arcsec\ $L_{X,34}^{1/2} \eta_{\rm
  X,-4}^{-1/2}f^{-1/2} p_{-10}^{-1/2} d_{4}^{-1}$, and a factor of a
few larger size of the X-ray nebula, comparable with the observed
nebula radius of $\sim150$\arcsec. This allows one to assume that the
detected nebula was created in a burst (or a series of bursts), which
is in line with our first assumption in Section 4.1, that likely the magnetar
experienced a bursting episode before obs.~1, which was not
directly detected.

We can in principle connect the nebula size (and even the softness of the spectrum) with synchrotron cooling. First of all, it is worth noting that the magnetic field at the shock (if there is a shock) does {\em not} depend on the neutron star surface magnetic field -- it is determined by the balance of the wind pressure and the ambient pressure and depends on the latter and the magnetization parameter $\sigma$ (i.e., the ratio of the electromagnetic energy flux to the kinetic energy flux): $B_s \sim [8\pi\sigma p/(1+\sigma)]^{1/2} \sim 50 [p_{-10}\sigma/(1+\sigma)]^{1/2}$ $\mu$G, upstream of the shock, and it can be a factor of 3 higher immediately downstream of the shock \citep{kennel84ApJ:mhdwind}.  This, in particular, means that the softness of the nebula spectrum is not due to a higher magnetic field in the nebula. The magnetization parameter $\sigma$ is, unfortunately, quite uncertain for the putative magnetar winds. It is believed to be $\ll 1$ for PWNe 
(e.g., $\sim 10^{-3}$ for the Crab), but it may be higher in magnetars. Therefore, the actual value of the magnetic field in the shocked magnetar
flow remains uncertain; it might be as low as a few $\mu$G (for small $\sigma$ and low-pressure ambient medium) or as high as a few mG (for large $\sigma$ and high-pressure medium). Therefore, we will simply scale the field as $B = 10^{-4} B_{-4}$ G.

The synchrotron cooling time for an electron with Lorentz factor
$\gamma$ can be estimated as $\tau_{\rm syn} = 5\times 10^8
\gamma^{-1} B^{-2}\, {\rm s} \sim 5\times 10^{8} \gamma_8^{-1}
B_{-4}^{-2}\, {\rm s} \sim 5\times 10^{8} B_{-4}^{-3/2}$ s, where for
synchrotron emission in the X-ray band we used $\gamma^2_8 B_{-4} \sim
(E/5\,{\rm keV}$).

The shocked wind flows from the magnetar with mildly relativistic
velocities (e.g., $c/3$ for an isotropic outflow -- see
\citealt{kennel84ApJ:mhdwind}). Multiplying $\tau_{\rm syn}$ by the
flow velocity, we obtain a distance from the magnetar where the X-ray
synchrotron radiation still can be observed: $R_{\rm MWN} \sim 5\times
10^{18} B_{-4}^{-3/2}$ cm, which corresponds to an angular distance of
$\sim 84\arcsec\ B_{-4}^{-3/2}$,  quite close to the observed size for
$B\sim 60$ $\mu$G. Thus, the observed size can be explained by the
synchrotron cooling of the outflowing electrons in a reasonable
magnetic field.

The cooling time also determines the lifetime of the putative MWN after
the end of the magnetar activity period. For instance, for $B\sim 60$
$\mu$G, $\tau_{\rm syn} \sim 30$ years, which means that the MWN can
be observable in X-rays around quiescent (even undetectable) magnetars
if these were in an active state years ago; it would also explain the
detection of the MWN in Obs.~1.

Finally, we would like to discuss some other possibilities for the
origin of the extended X-ray emission around \src. The source lies in
the center of a crowded field filled with many other high energy
sources. It lies almost at the center of the extended TeV source
HESS~J1834$-$087 \citep{aharonian06ApJ:HESSsur1}, and within the SNR
W41 (K+12) and a dense GMC \citep{tian07ApJ:1834}. The high absorbing
column density toward \src\ is most likely related to the GMC, which in
turn is causing the scattering halo emission. An
anisotropic dust distribution within the GMC could cause an
asymmetrical halo emission, leading to region A and region B emanating
from the same region and having the same physical origin. To test this
hypothesis we extracted the spectrum of region~A+region~B during
obs.~2 and fit it with an absorbed power-law. We find a hydrogen
column density $N_{\rm H}=17_{-3}^{+4}\times10^{22}$~cm$^{-2}$,
consistent with the point source absorbing column, and a power-law photon
index $\Gamma=3.4\pm0.5$, harder than the point source spectrum, indicating
that a halo interpretation for region~A+region~B is unlikely. Hence,
the nature of these two regions is indeed different as indicated by
their different spectral properties (Section~\ref{sec:spec}). Moreover,
the detection of region~B during obs.~1, when the source was in
quiescence, poses a challenge to such an interpretation. Another
possibility for the region B emission could be some contribution from
the SNR W41, in the form of either thermal emission from shocked gas
or non-thermal synchrotron emission \citep[see][for a
review]{vink12AARv:SNR}. However, the fluxes of both region~A and
region~B varied with the source flux, implying a tight connection
between the two and the SGR. Deeper high-resolution multiwavelength
observations would be of great value to better understand the physical
properties and emission processes of the \src\ putative MWN, and would
help shed light on the connections between the many point-like and
extended sources existing in this crowded field.

\section*{Acknowledgments}
 
This work is based on observations with \xmm\, an ESA science mission
with instruments and contributions directly funded by ESA Member
States and the USA (NASA). The work by OYK and GGP was partly
supported by NASA grants NNX09AC81G and NNX09AC84G, NSF grants
AST09-08733 and AST09-08611, and by the Ministry of Education and
Science of the Russian Federation (contract 11.G34.31.0001). The
authors are grateful to Norbert  Schartel for his decision to allocate
\xmm\ TOO time  for  observation of \src. The authors would like to
thank the referee for the constructive comments that helped improve
the quality of the manuscript.

\end{document}